\documentclass[conference]{IEEEtran}
\IEEEoverridecommandlockouts
\usepackage{algorithm, setspace}
\usepackage{comment}
\usepackage{mathtools}
\usepackage{footnote}
\usepackage{tabularx}
\usepackage{booktabs}
\usepackage[noend]{algpseudocode}
\usepackage{multicol}
\usepackage{multirow}
\usepackage{subcaption}
\usepackage{cleveref}
\usepackage{xcolor}
\usepackage{url}
\floatname{algorithm}{Algorithm}

\newtheorem{definition}{Definition}

\DeclarePairedDelimiter\floor{\lfloor}{\rfloor}

\begin{document}

\title{Xling: A Learned Filter Framework for Accelerating High-Dimensional Approximate Similarity Join}

\author{\IEEEauthorblockN{1\textsuperscript{st} Yifan Wang}
\IEEEauthorblockA{
\textit{University of Florida}\\
wangyifan@ufl.edu}
\and
\IEEEauthorblockN{2\textsuperscript{nd} Vyom Pathak}
\IEEEauthorblockA{
\textit{University of Florida}\\
v.pathak@ufl.edu}
\and
\IEEEauthorblockN{3\textsuperscript{rd} Daisy Zhe Wang}
\IEEEauthorblockA{
\textit{University of Florida}\\
daisyw@ufl.edu}
}

\maketitle

\begin{abstract}
Similarity join is a critical and widely used operation in multi-dimensional data applications, which finds all pairs of close points  within a given distance threshold. Being studied for decades, many similarity join methods have been proposed, but they are usually not efficient on high-dimensional space due to the curse of dimensionality and data-unawareness. 
Inspired by the Bloom join in RDBMS, we investigate the possibility of using metric space Bloom filter (MSBF), a family of data structures checking if a query point has neighbors in a multi-dimensional space, to speed up similarity join. However, there are several challenges when applying MSBF to similarity join, including excessive information loss, data-unawareness and hard constraint on the distance metric, because of which few works are designed in this way.

In this paper, we propose Xling, a generic framework to build a learning-based metric space filter with any existing regression model, aiming at accurately predicting whether a query point has enough number of neighbors. The framework provides a suite of optimization strategies to further improve the prediction quality based on the learning model, which has demonstrated significantly higher prediction quality than existing MSBF.  
We also propose XJoin, one of the first filter-based similarity join methods, based on Xling. By predicting and skipping those queries without enough neighbors, XJoin can effectively reduce unnecessary neighbor searching and therefore it achieves a remarkable acceleration. Benefiting from the generalization capability of deep learning models, XJoin can be easily transferred onto new dataset (in similar distribution) without re-training.  
Furthermore, Xling is not limited to being applied in XJoin, instead, it acts as a flexible plugin that can be inserted to any loop-based similarity join methods for a speedup. Our evaluation shows that Xling not only leads to the high performance of XJoin (e.g., being up to 17x faster than the baselines while maintaining a high quality), but also be able to further speed up many existing similarity join methods with quality guarantee.            
\end{abstract}


\section{Introduction}
\label{sec:intro}
In multi-dimensional data management, metric space range search (shortly called \textit{range search}) and similarity join are critical operators. 
Given a distance threshold $\epsilon$, the former operation finds all points in a dataset $D$ whose distance to a given query point is less than $\epsilon$, while the latter operation finds all pairs of points between two datasets $R$ and $S$ whose distance is less than $\epsilon$. As range search can be seen as a special case of similarity join where $|R| = 1$ or $|S| = 1$, unless it is necessary, we will only discuss similarity join in this paper.  
With the emergence of deep learning, high-dimensional neural embedding 
has been widely adopted as the data representation in a wide range of applications, which raises the demands for effective and efficient similarity join over high-dimensional data. Those applications include near-duplicate detection \cite{near-duplicate-detect-app-1-paper-sim-join-with-vector, near-duplicate-detect-app-2-place-sim-join-with-vector, near-duplicate-detect-app-3-video-sim-join-with-vector, lsh-sim-join-near-dup-detect-1, lsh-sim-join-near-dup-detect-2}, data integration \cite{data-integration-app-1-with-vector, er-tasks-sim-join-1-sim-threshold-with-vector, er-tasks-sim-join-4-most-typical-sim-threshold-with-vector}, data exploration \cite{table-discovery-by-embedding-cosine-similarity-0.7, table-discovery-by-embedding-cosine-similarity-0.8, table-discovery-app-1-dong2021efficient}, privacy \cite{lsh-sim-join-privacy-1, lsh-sim-join-privacy-2} and so on. 
Specifically, distance between embeddings reflects the semantic similarity between the data objects, meaning that the applications have to frequently search for close points in the embedding space, i.e., the similarity join. 
In many real-world use cases, the similarity join 
is approximate, i.e., they do not require 100\% accurate results, but do demand high processing speed, especially on large-scale data. A typical example is near-duplicate video retrieval \cite{near-duplicate-detect-app-3-video-sim-join-with-vector} that identifies online videos with identical or almost identical content during the search process to diversify the video search results, in which case missing a few pairs of similar videos is acceptable, while fast response is required.

Existing work on efficient similarity join mainly includes two categories: space-grid~\cite{ego-based-hilbert-join, ego-join, ego-star-join, super-ego} and locality-sensitive hashing (LSH) based methods~\cite{lsh-for-sim-join, lsh-sim-join-c2net-distributed-system, lsh-sim-join-near-dup-detect-1, lsh-sim-join-near-dup-detect-2, lsh-sim-join-privacy-1, lsh-sim-join-privacy-2}. The former splits the data space into grids and joins the points within the same or neighboring grids, while the latter is essentially an adoption of RDBMS hash-join principle onto the high-dimensional join, i.e., hashing one dataset by LSH and then probing it for the points in another dataset. However, the grid-based methods poorly perform in very high-dimensional space due to the curse of dimensionality, while the unawareness for data distribution usually causes a non-trivial performance degradation to LSH-based methods on unevenly distributed data \cite{hierarchical-non-uniform-lsh}. In addition, the grid-based methods are usually exact while LSH-based methods are approximate, i.e., their accuracies are respectively 100\% and lower than 100\%. In this paper we focus on approximate similarity join.

There is one more possible way that speeds up similarity join with \textit{metric space Bloom filter}, inspired by the Bloom join in RDBMS (which uses a Bloom filter to prune unnecessary probings).
Metric space Bloom filter (MSBF) is designed for checking whether a query point has neighbors within the given distance threshold. Among the various MSBFs~\cite{locality-sensitive-bloom-filter, lsbf-variant-1, lsbf-variant-2-no-fpr, lsbf-variant-3-multi-gran-hamming, lsbf-variant-4-multi-gran, dsbf, locality-sensitive-bloom-filter-2019}, Locality-Sensitive Bloom Filter (LSBF)~\cite{locality-sensitive-bloom-filter} is the most widely used, based on which a substantial number of MSBFs have been developed. Mirroring to the hashing functions in Bloom filter, LSBF utilizes LSH functions to map a multi-dimensional point to several bits in the bit array, and determines the neighbor existence by counting the non-zero bits.   
Given datasets \textit{R} and \textit{S}, an LSBF can be built on \textit{R} and each point $s \in S$ will act as a query. By skipping the range search for those negative queries (i.e., the queries having no neighbors in $R$), the similarity join can be accelerated.
Note that unlike Bloom filter, most MSBFs cannot guarantee a zero false negative rate since they are based on LSH which raises both false positives and negatives.    

According to our evaluation, negative queries usually take up a non-trivial portion (20\% $\sim$ 90\%), meaning that the filtering should result in a significant performance improvement. But few methods are designed in such a way, due to several problems of MSBF:
(1) Given the fact that they are built on top of LSH, their effectiveness is also limited by the data-unawareness.
(2) They lose more information compared to the original LSH because of further mapping LSH values to one-dimension bit array, which additionally lowers their effectiveness.  
(3) They indicate the index of target bit by the LSH value, and this disables them to support many popular distance metrics, like cosine distance where the LSH values are always 0 or 1.

Inspired by the ``learned Bloom filter''~\cite{case-for-learned-index-structures-kraska2018case} that enhances Bloom filter with machine learning, we propose a new type of MSBF based on machine learning techniques which addresses the problems above. 
Specifically, in this paper, we propose \textbf{Xling}, a generic framework for building a MSBF with deep learning model. Instead of LSH, Xling relies on the  \textit{learned cardinality estimation} techniques which utilize regression models to predict the number of neighbors for range search before actually executing it.  
Xling is designed to be generic such that any cardinality estimator (or simply, regression model) can be encapsulated into an effective MSBF. 
Its core is a cardinality estimator with a \textit{Xling decision threshold} (XDT). For a range query, Xling predicts the result cardinality, then determines the query is positive (i.e., having enough neighbors) if the prediction exceeds XDT, otherwise it is negative (i.e., being without enough neighbors). Note that we mention ``enough neighbors'' instead of ``any neighbors'', which reveals an advanced feature: Xling can determine whether the query point has more than $\tau$ neighbors, where $\tau$ is a user-determined number. And Xling downgrades to a general MSBF when $\tau = 0$. We call such a feature ``filtering-by-counting''.

By learning the data distribution, Xling solves the data-unawareness problem, and  essentially as a regressor, it is not limited to any specific distance metric. 
Note that we have mentioned three thresholds until now, the \textit{distance threshold} $\epsilon$, the \textit{Xling decision threshold} XDT, and the \textit{neighbor threshold} $\tau$. To make it clearer, we will indicate them respectively with ``distance threshold'', ``decision threshold'' and ``neighbor threshold'', or directly using their symbols.

Xling deploys novel optimization strategies to further improve the performance, including strategy to select $\epsilon$ values for effective training, and strategy to select the best XDT. 
Furthermore, Xling can be applied as a plugin onto many existing similarity join methods to significantly enhance their efficiency with tiny or no quality loss. 
We have applied Xling to a brute-force nested-loop similarity join, named \textbf{XJoin}, which achieves significantly higher efficiency (up to 14.6x faster) than the existing high-performance similarity join methods (some are used in industry), with a guarantee of high quality. In these applications, filtering-by-counting enables Xling to help the base similarity join method ignore the queries having only a trivial number of neighbors (e.g., 3 or 5, or even 50), which makes the acceleration more significant with tiny sacrifice of recall.       
In addition, we also apply Xling onto those prior methods and show that they are substantially accelerated with slightly more quality loss. Finally, XJoin and Xling are evaluated on their generalization capability, and the results prove that they can be transferred to updated or fully new dataset without re-training the learning model.       
To our best knowledge, We are the first to propose such a learning-based MSBF and XJoin is among the first practical filter-based similarity join methods for high-dimensional data.

The main contributions of this paper are as follows:
\begin{enumerate}
    \item We propose XJoin, the first filter-based similarity join method for high-dimensional data, which is both efficient and effective.  
    \item We propose Xling, a generic framework for constructing learned metric space Bloom filters with general regression models. 
    \item We design performance optimization strategies in Xling, including selection of $\epsilon$ values for effective training and adaptive computing of XDT.
    \item We conduct extensive evaluation to show the efficiency, effectiveness, usefulness and generalization of Xling and XJoin, as well as the remarkable performance improvement by applying Xling to other similarity join methods.
\end{enumerate}
The rest of this paper is organized as follows: Section \ref{sec:related-work} introduces the prior works related to the techniques in this paper. Section \ref{sec:preliminaries} formally defines the key problems studied by this paper and the important notations being used. Section \ref{sec:arch} presents the architecture of Xling and the workflow of applying Xling to enhancing similarity join. Section \ref{sec:optim-strategy} discusses details about the optimization strategies integrated in Xling. And Section \ref{sec:exps} reports and analyzes the evaluation results.

\section{Related work}
\label{sec:related-work}
\noindent \textbf{Learned Bloom filter: } The learned Bloom filter is first proposed by \cite{case-for-learned-index-structures-kraska2018case} which treats the existence checking in Bloom filter as a classification task, thereby replaces the traditional Bloom filter with a deep learning based binary classifier followed by an overflow Bloom filter (to double check the negative outputs of the classifier to guarantee a zero false negative rate). 
And there have been many following works that further improve the learned Bloom filter \cite{learned-bloom-filter-multi-dim, sandwiching-learned-bf, learned-bloom-filter-improve-hash-func} by adding auxiliary components or improving the performance of the hashing functions being used.

\noindent \textbf{Learned cardinality estimation: } Learning models have been widely used to predict the number of neighbors for a metric space range search, which is called ``learned cardinality estimation''. The learned cardinality estimation techniques treat the task as a regression problem and solve it using regression models.
The state-of-the-art methods~\cite{card-est-sun2021learned, cardnet-wang2020monotonic, tutorial-card-est-high-dim, selnet-wang2021consistent} are usually based on deep regression models to effectively learn the data distribution and make more accurate prediction than the non-deep approaches. Recursive Model Index(RMI) \cite{rmi-kraska2018case} is a hierarchical learning model that consists of multiple sub-models, where each sub-model is a regression model like neural network. CardNet \cite{cardnet-wang2020monotonic} consists of a feature extraction model and a regression model. The raw data is transformed by the feature extraction model into Hamming space, which will then used by the regression model to predict its cardinality. SelNet \cite{selnet-wang2021consistent} predicts the cardinality by a learned query-dependent piece-wise linear function.

\noindent \textbf{Metric space Bloom filter: } 
DSBF~\cite{dsbf} and LSBF~\cite{locality-sensitive-bloom-filter} are two of the representatives for metric space Bloom filter (MSBF), and following them many MSBF variants and relevant applications have been developed. \cite{lsbf-variant-4-multi-gran} enables LSBF to handle multiple Euclidean distance granularity without re-building the data structure. \cite{lsbf-variant-3-multi-gran-hamming} extends \cite{lsbf-variant-4-multi-gran} to Hamming distance. \cite{lsbf-variant-2-no-fpr} proposes a DSBF variant with zero false negative rate theoretically. 
Since most of those works are built on top of LSH, their effectiveness is usually limited by unknowing of the data distribution.     

\noindent \textbf{Similarity join: } This problem can be further classified into two sub-types, exact and approximate similarity join, where the former is to exactly find all the truly close points while the latter allows some errors.

One family of the state-of-the-art efficient methods for exact similarity join is the epsilon-grid-order (EGO) based methods, including EGO-join \cite{ego-join}, EGO-star-join \cite{ego-star-join}, Super-EGO \cite{super-ego}, FGF-Hilbert join \cite{ego-based-hilbert-join}, etc. They work by splitting the space into cells and sorting the data points along with those cells, which will then help reduce unnecessary computation in the similarity join.   

Approximate similarity join methods are usually based on LSH \cite{lsh-for-sim-join, sim-join-using-hashing, lsh-sim-join-c2net-distributed-system, lsh-sim-join-near-dup-detect-1, lsh-sim-join-near-dup-detect-2, lsh-sim-join-privacy-1, lsh-sim-join-privacy-2}. 
The problem of those methods is data unawareness, i.e., the hashing-based space partitioning usually does not consider the data distribution, which may leads to imbalanced partitions that significantly lower the overall search performance \cite{hierarchical-non-uniform-lsh}.

\section{Preliminaries}
\label{sec:preliminaries}
\begin{table}[!h]
\small
  \begin{tabularx}{\columnwidth}{lX}
    \toprule
     Notation & \multicolumn{1}{c}{Description}  \\
    \midrule
    \ $\epsilon$ & The distance threshold for range search and similarity join  \\
    \midrule
    \ $\tau$ & The neighbor threshold, i.e., whether or not a query has more than $\tau$ neighbors \\
    \midrule
    \ \textit{XDT} & The Xling decision threshold to classify the prediction as positive or negative \\
    \midrule
    \ $N$ & Dataset size  \\
    \midrule
    \ $|R|$, $|S|$ & The sizes of set $R$ and $S$ \\
    \midrule
    \ $MAE$, $MSE$ & Mean absolute error and mean squared error \\
    \midrule
    \ $FPR$, $FNR$ & False positive rate and false negative rate \\
    \bottomrule
  \end{tabularx}
  \vspace{1mm}
  \caption{List of notations used in this and following sections}
  \label{tab:notations}
  \vspace{-5mm}
\end{table}

This section presents the critical notations frequently used in this paper and defines the key problems being studied, i.e., range search and similarity join. The notations are listed and explained in Table~\ref{tab:notations}, and more details of them are introduced where they are first referred in this paper.

In multi-dimensional space, range search is defined as the search of all data points whose distance to the query is smaller than a given threshold under some distance metric. And similarity join is to find all close point pairs whose distance is less than a threshold between two datasets. The formal definitions are as below:

\begin{definition}[range search]
\label{def:range-search}
    Given a dataset $P = \{p_i|i = 1, 2, ..., n\}$, a distance metric $d(\cdot, \cdot)$, a query point $q$ and a radius/threshold $\epsilon$, range search tries to find a set of data points $P^\ast$ such that for any $p^\ast_i \in P^\ast$, $d(q, p^\ast_i) \leq \epsilon$.
\end{definition}

\begin{definition}[similarity join]
\label{def:sim-join}
   Given two datasets $R$ and $S$, a distance threshold $\epsilon$, and a distance metric $d(\cdot, \cdot)$, the similarity join between $R$ and $S$ is denoted by $R \bowtie_\epsilon S$, which combines each point $r \in R$ with each point $s \in S$ that is close/similar enough to $r$ (i.e., with distance smaller than or equal to $\epsilon$). Formally
    \begin{align}
        R \bowtie_\epsilon S = \{(r, s) | \forall r \in R, \forall s \in S\ where\ d(r,\ s) \leq \epsilon\}
    \label{eq:sim-join}
    \end{align}
\end{definition}

There are three critical ``thresholds'' in this paper, as listed in Table \ref{tab:notations}, the ``distance threshold'' $\epsilon$, the ``neighbor threshold'' $\tau$ and the ``Xling decision threshold'' XDT. To avoid confusion, we further explain them here: 
\begin{enumerate}
    \item The distance threshold $\epsilon$ is part of the range search and similarity join operations, which indicates the search range as shown in Definition \ref{def:range-search} and \ref{def:sim-join}.
    \item $\tau$ is a threshold for the \textit{groundtruth} neighbors. If a query truly has more than $\tau$ neighbors, we call it a \textit{groundtruth positive}, otherwise it is a \textit{groundtruth negative}.     
    \item The Xling decision threshold (XDT) is the threshold to classify the query based on the \textbf{predicted} neighbors. If a query is predicted as having more than XDT neighbors, we name it as a \textit{predicted positive}, otherwise it is predicted negative.      
\end{enumerate}
Note that we do not directly use $\tau$ to threshold the predictions. This is because different cardinality estimation models have different prediction accuracy, leading to different predicted values for the same query. So it is necessary to use an adpative threshold driven by both model and data to classify the predictions, which is XDT, such that we can control the false positive or negative rate. More details are introduced in Section~\ref{sec:train-eps-sel}.

\section{Architecture and workflow}
\label{sec:arch}
\begin{figure}[!h]
  \centering
  \includegraphics[width=0.7\columnwidth]{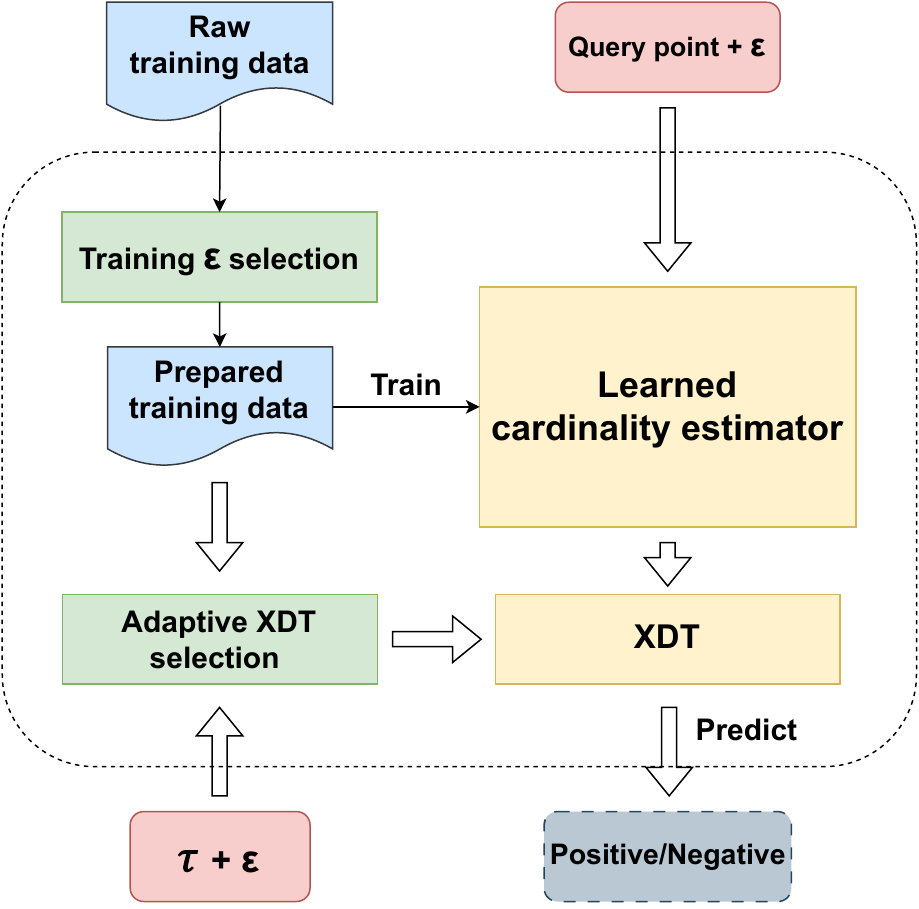}
  \caption{Architecture and workflow of Xling}
  \setlength{\belowcaptionskip}{-50pt}
  \label{fig:arch}
  \vspace{-4mm}
\end{figure}

\subsection{The core and optimization strategies}
Figure \ref{fig:arch} illustrates the overall architecture and workflow of Xling. The core is the learned cardinality estimator and XDT (yellow boxes). The green boxes represent the optimization strategies deployed in the training and predicting stages. The blue shapes stand for the training data, including the raw data (which is input from the external) and the prepared data (which is generated based on the raw data inside Xling). The workflow of Xling mainly relies on the prepared training data instead of the raw data. In addition, the pink boxes indicate the query-time inputs, including the query point, $\epsilon$ and $\tau$, where the query point and $\epsilon$ are fed into the learned cardinality estimator while the $\epsilon$ and $\tau$ are used to compute XDT. The grey box stands for the output, i.e., whether the query point is positive (with enough neighbors) or negative (having insufficient neighbors). For the workflow, the solid arrows present the offline training workflow, while the hollow arrows indicate the online workflow for querying and XDT computing.

The overall architecture includes (1) the core learned cardinility estimator and XDT and (2) the surrounding optimization strategy modules. \cite{case-for-learned-index-structures-kraska2018case} uses a classifier and a decision threshold to build a learned Bloom filter, while we use a regressor and the XDT to construct Xling, which is similar to the learned accelerator framework (LAF) in \cite{laf}. 
Specifically, LAF proves that learned cardinality estimator with a proper decision threshold can significantly accelerate range-search based high-dimensional clustering. As similarity join is also based on range search, such a solution works for it too. And unlike the case of learned Bloom filter, since filtering-by-counting requires estimating the specific number (mentioned in Section~\ref{sec:intro}), regressor is the best choice rather than classifier. 
But in LAF, the decision threshold has to be determined by grid search. To overcome this shortcoming, we design an \textit{adaptive XDT selection} strategy based on the training data such that XDT can be efficiently computed. More details are introduced in Section \ref{sec:xdt-sel}.

To reduce the false results, in addition to adaptively selecting a proper XDT, we also propose an adaptive \textit{training $\epsilon$ selection} strategy to sample the most representative $\epsilon$ values for training the cardinality estimator effectively and thereby reducing the prediction error. The existing learned cardinality estimation studies (e.g., \cite{mono-card-est-cardnet}) usually select the training $\epsilon$ values uniformly, which is not optimal, and we show in this paper that our proposed strategy generates a more representative training set and leads to a more effective model training. This training $\epsilon$ selection strategy is further discussed in Section \ref{sec:train-eps-sel}. 

\subsection{Training and querying workflows}
There are two workflows in Xling, the offline training workflow and online querying workflow. In Figure \ref{fig:arch}, the solid arrows illustrate the training workflow. As introduced in Definition \ref{def:sim-join} and Section \ref{sec:intro}, the two point sets to be joined are denoted by $R$ and $S$, and without loss of generality, we assume that the size of $R$ ($|R|$) is larger than size of $S$ ($|S|$). Then $R$ is used as the training set (i.e., the ``raw training data'' in Figure \ref{fig:arch}) to train Xling while $S$ acts as the queries. The raw training data includes all points in $R$ without $\epsilon$ information.

\subsubsection{Training}
\label{sec:training-workflow}
The first step in the training workflow is concatenating the selected $\epsilon$ values onto each training point $r_i \in R$ (where $i=1, 2, ..., |R|$), generating the ``prepared training data'' that is a Cartesian product between the point set $R$ and $\epsilon$ set $\mathcal{E}$, i.e., the prepared training data is the set \{($r_i$, $\epsilon_j$) $|$ $\forall (r_i, \epsilon_j) \in R \times \mathcal{E} $\}, where $r_i$ is a multi-dimensional vector and $\epsilon_j$ is a real number. Here the set $\mathcal{E}$ is generated by the training $\epsilon$ selection strategy (Section \ref{sec:train-eps-sel}).

To make it clearer, suppose $R$ is a toy raw training set of 2 points $r_a$ and $r_b$, and $\mathcal{E}$ includes two selected values $\epsilon_1$ and $\epsilon_2$. Then the prepared training set looks like \{($r_a$, $\epsilon_1$), ($r_a$, $\epsilon_2$), ($r_b$, $\epsilon_1$), ($r_b$, $\epsilon_2$)\} associated with their targets (i.e., the groundtruth numbers of neighbors).
Then the learned cardinality estimator is trained with the prepared training data. 

\subsubsection{Querying}
The querying workflow is presented with hollow arrows in Figure \ref{fig:arch}. First, $\epsilon$ and $\tau$ are input to compute the XDT using the adaptive XDT selection strategy based on the prepared training data. Second, the query point and the $\epsilon$ are concatenated and fed to the cardinality estimator to predict the number of neighbors. Finally, the prediction is compared with XDT and the answer (positive or negative) is determined. 

\subsection{Use in similarity join}
\label{sec:use-in-simjoin}
Use of Xling to speed up similarity join is straightforward: given $\epsilon$, $\tau$, $R$ and $S$, supposing $|R| > |S|$, Xling is first trained on $R$, then for each query point $s \in S$, Xling predicts whether $s$ has enough neighbors in $R$ under the $\epsilon$. If yes, the range search (either a brute-force search or an indexed search like using LSH) will be executed in $R$ for query $s$, otherwise $s$ is predicted as negative and the search for it will be skipped. In this way, unnecessary range search is reduced and the join efficiency is improved.

As a generic filter, Xling can be applied onto any nested-loop based similarity join algorithms to speed them up. When using deep regressor as the core cardinality estimator in Xling, its prediction time complexity for each query is constant (O(1)) with the data scale, which in practice can be further accelerated by GPU. And the training time is not an issue due to the generalization capability of the deep models, i.e., a trained estimator can be used on any other dataset with similar distribution, which is proved by our evaluation in Section~\ref{sec:exp-generalization}.

We implement XJoin by applying Xling onto a naive nested-loop similarity join method which does a brute-force range search for each query. Our evaluation presents that XJoin outperforms the state-of-the-art similarity join baselines. We also apply Xling to some approximate similarity join methods, which shows Xling successfully improves their speed-quality trade-off capability.

\section{Optimization strategies}
\label{sec:optim-strategy}

In this section we introduce the details for the optimization strategies we propose in Xling. 

\subsection{Training $\epsilon$ selection}
\label{sec:train-eps-sel}
\cite{mono-card-est-cardnet} uniformly samples the training $\epsilon$ from range [0, $\theta_{max}$], where $\theta_{max}$ is a user-determined upper limit for the distances. However, such a selection strategy is not optimal. We show in this section that the cardinality estimation models can be trained more effectively by using our data-aware threshold selection strategy. Furthermore, our strategy can be generalized to better solve a family of regression problems.  

The specific family of regression problems is named \textit{Condition-based Regression} (shortly denoted by \textit{CR}) and defined as follows:
\begin{definition}[Condition-based Regression problem]
Given a dataset $\mathcal{D}$, a condition $c$ associated with each data point $p \in \mathcal{D}$, and a target $t$ corresponding to each $(p, c)$ pairs, the CR problem is to learn the relationship between the presented $(p, c)$ and $t$, then predict the corresponding $t$ for any given $(p, c)$.    
\end{definition}

In the context of learned cardinality estimation, the condition is the $\epsilon$ while the target is the number of neighbors.

In most Condition-based Regression problems, the condition is usually a  continuous variable (like distance) whose values cannot be fully enumerated, therefore it is worth studying how to sample the conditions for better training the regression model in order to lower prediction error, which is formally defined as such:

\begin{definition}[Training condition selection for CR]
\label{def:train-cond-sel}
Given a dataset $\mathcal{D}$ of size $n$, a set of $m$ candidate values \{$c_{i1}, c_{i2}, ..., c_{im}$\} of the condition $c_i$ for each data point $p_i \in \mathcal{D}, 1 \leq i \leq n$, the corresponding target $t_{ij}$ for each $(p_i, c_{ij}), 1 \leq j \leq m$, and a sampling number $s$, the training condition selection task is to select $s$ pairs from the $m$ $(c_{ij}, t_{ij})$ pairs to form $s$ training tuples for each point $p_i$, which results in totally $sn$ training tuples for the whole dataset $\mathcal{D}$, such that the mean regression error is minimized. 
\end{definition}

Note that our discussion starts from a discrete candidate set, instead of directly beginning from the continuous range (like [0, $\theta_{max}$]). This is just to unify the discussion between continuous-range uniform sampling and our sampling method which requires preprocessing to discretize the range, which will not lose generality. 
$m$ is usually large (such that the values are dense enough) to approximate the case of the continuous condition values, so we do not use all the $m$ candidate values to generate $mn$ training tuples, as the memory space is limited. For example, in our evaluation, $m = 100$, in which case $mn$ tuples require more memory space than that of our evaluation machine. Therefore only the $s$ sampled conditions are used to form the training tuples.

Uniformly sampling training conditions cannot reflect the unevenly data distribution in real-world datasets. Therefore, we design a generic adaptive strategy to further fine-tune the initial uniformly sampling conditions for CR problem based on the density of the targets, such that the resulting training tuples are more representative for the data distribution of the whole dataset. Our adaptive training condition selection (ATCS) strategy (Algorithm \ref{alg:train-cond-sel}) follows the steps below for each data point $p$ in the training set:
\begin{enumerate}
    \item Given the uniformly sampled candidate conditions $\mathcal{C}_p$ and the corresponding targets $\mathcal{T}_p$ for point $p$ (i.e., when condition $c_i \in \mathcal{C}_p$ is applied to point $p$, the target is correspondingly $t_i \in \mathcal{T}_p$), the minimum and maximum targets are found ($t_{min}$ and $t_{max}$, line 5) and the interval [$t_{min}, t_{max}$] is then partitioned into $s$ bins evenly (line 6).
    \item Then each candidate condition is mapped into one bin based on its paired target (line 7-8).
    \item Finally the specific number of condition-target pairs are sampled from each bin (line 10-11) according to the fraction of the bin size ($|\mathcal{B}_i|$) over total number of the conditions ($|\mathcal{C}_p|$), where bin size is the number of pairs in that specific bin.
    \item Since some bins may generate zero samples (i.e., when $s|\mathcal{B}_i| < |\mathcal{C}_p|$), the final sampled condition-target pairs may be not enough given the required $s$. In such a case, the rest\protect\footnote{In our evaluation, the rest of the pairs (which are randomly chosen) usually occupy 10\% $\sim$ 20\% in the final training set for the whole $\mathcal{D}$, i.e., $0.1s|\mathcal{D}| \sim 0.2s|\mathcal{D}|$} of the pairs will be randomly chosen from the unselected ones that are not yet included in the samples above (line 12-13). 
\end{enumerate}
The final training set for the whole $\mathcal{D}$ is the collection of the training tuples ($p$, $c$, $t$) generated by the strategy for each $p \in \mathcal{D}$, i.e., the final training set includes $s|\mathcal{D}|$ tuples.

\begin{algorithm}
\setstretch{0.8}
\small
\caption{Adaptive training condition selection (ATCS) strategy}
\label{alg:train-cond-sel}
\begin{algorithmic}[1]
\Require{Dataset $\mathcal{D}$, map from each data point to its uniformly sampled candidate condition list $\mathcal{C}$, map from each point to its target list $\mathcal{T}$, sampling number $s$} 
\Ensure{set of resulting training tuples $\mathcal{R}$}
\State $\mathcal{R}$ := $\emptyset$
\For{\textbf{each} point $p$ in $\mathcal{D}$} 
    \State $\mathcal{C}_p$ := $\mathcal{C}(p)$  \Comment{the condition list for $p$ } 
    \State $\mathcal{T}_p$ := $\mathcal{T}(p)$  \Comment{the target list for $p$, one-to-one corresponding to $\mathcal{C}_p$}
    \State $t_{min}$, $t_{max}$ := min($\mathcal{T}_p$), max($\mathcal{T}_p$)
    \State Split interval [$t_{min}$, $t_{max}$] into $s$ bins evenly
    \For{\textbf{each} $c$ in $\mathcal{C}_p$ and corresponding $t$ in $\mathcal{T}_p$}
        \State Place $(c, t)$ into a bin bounded by $[t_a, t_b)$ such that $t \in [t_a, t_b)$ 
    \EndFor
    \State $\mathcal{S}$ := $\emptyset$ \Comment{the selected $(c,t)$ collection for $p$}
    \For{\textbf{each} bin $\mathcal{B}_i$}
        \State $\mathcal{S}$ := $\mathcal{S} \cup \{\floor*{\frac{s|\mathcal{B}_i|}{|\mathcal{C}_p|}}$ randomly sampled $(c, t)$ pairs from $\mathcal{B}_i\}$ 
    \EndFor
    \If{$|\mathcal{S}| < s$}
        \State Fill with random samples from unselected $(c,t)$ until $|\mathcal{S}| =s$  
    \EndIf
    \State $\mathcal{R}$ := $\mathcal{R} \cup \{(p,c_s,t_s) | (c_s,t_s) \in \mathcal{S}\}$ \Comment{combine all pairs in $\mathcal{S}$ with $p$}
\EndFor
\State \textbf{return} $\mathcal{R}$
\end{algorithmic}
\end{algorithm}

Unlike uniformly sampling, ATCS is data-aware by binning the condition-target pairs to estimate the density of the targets and then sampling the final conditions based on the density, which generates more representative training conditions and targets. 
For example, let $m = 100, s = 5$, i.e., the interval of targets [$t_{min}$, $t_{max}$] is split into 5 bins evenly and their corresponding condition values are placed into the bins accordingly. Supposing the numbers of conditions from $\mathcal{B}_1$ to $\mathcal{B}_5$ are 59, 1, 19, 1, 20, the uniformly sampling will select 2, 1, 0, 1, 1 conditions from them (i.e., selecting one per 20), while ATCS will first select 2, 0, 0, 0, 1 from them (Algorithm~\ref{alg:train-cond-sel} Line 11) then probably select the rest 2 conditions from $\mathcal{B}_1$, $\mathcal{B}_3$ and $\mathcal{B}_5$ which are the most dense areas (Algorithm~\ref{alg:train-cond-sel} Line 13). In this example, the uniformly sampling gets 2 out of the 5 sampled conditions from very sparse areas ($\mathcal{B}_2$ and $\mathcal{B}_4$) in the distribution of targets, which cannot well reflect the overall distribution, while ATCS probably selects all the 5 from dense areas and results in more representative training samples, by which the model can be trained more effectively. Our evaluation in Section \ref{sec:optim-strat-eval} shows that ATCS helps reduce 50\% $\sim$ 98\% of the prediction error (MAE and MSE).    

\subsection{Xling decision threshold (XDT) selection}
\label{sec:xdt-sel}
Xling decision threshold (XDT) determines whether the prediction means positive. Its value is influenced by $\tau$, the way it is computed, and the way to identify the groundtruth negative training samples, which will be introduced in this section. Note that XDT is determined purely based on the training set offline, regardless of the online queries, i.e., the computation of XDT does not raise overhead on the online querying.

Basically, XDT is computed using the groundtruth negative training samples  (i.e., the training points with no more than $\tau$ neighbors). 
We propose two ways to compute XDT: false positive rate (FPR) based and mean based.  
\begin{enumerate}
    \item FPR-based XDT selection: similarly to \cite{rmi-kraska2018case, multidim-learned-bloom-filter-based-on-classifier-tim, sandwiching-learned-bf}, given a FPR tolerance value $t_{fpr}$ (e.g., 5\%), this method lets the estimator make predictions for the training points and sets XDT such that the resulting filter FPR on training set is lower than $t_{fpr}$. 
    \item mean-based XDT selection: this method sets XDT to be the mean predicted value for all the groundtruth negative training samples. 
\end{enumerate}
In our evaluation (Section~\ref{sec:xdt-sel-eval}), FPR-based method usually results in a higher XDT, leading to more speedup and lower quality in end-to-end similarity join, so we design the mean-based method to provide the second option which results in a lower XDT and leads to higher quality and less speedup. They are useful in different situations as shown in our evaluation. In addition, no matter using FPR-based or mean-based selection, there is a trend that a larger $\tau$ will result in a higher XDT, which is straightforward to understand: a larger $\tau$ causes training samples with more neighbors to be negative, therefore cardinalities of the negative samples increase overall, making the computed XDT increased.

But both methods have a problem: we have to first identify the groundtruth negative samples in the training set, which is costly in the high-dimensional cases. Since the negative samples depend on $\epsilon$, and the training set only include a tiny portion of all possible $\epsilon$, the queried $\epsilon$ will probably not exist in training set (named ``out-of-domain $\epsilon$''), in which case intensive range search has to be executed to compute the negative samples from scratch.     

Therefore, it is non-trivial to study how to easily get the training targets (i.e., groundtruth cardinalities) under the out-of-domain $\epsilon$ such that the groundtruth negative samples can be identified without doing range search for each training point.
We propose an interpolation-based strategy to generate the approximate targets. Specifically, given a point, its $\epsilon$-cardinality curve is monotonically non-decreasing, i.e., with $\epsilon$ increasing, the cardinality will never decrease. So we can approximate the curve segment between two neighboring training $\epsilon$ values (denoted by $\epsilon_{1}$ and $\epsilon_{2}$) as linear, and use linear-interpolation to estimate the groundtruth cardinality for any training point under a out-of-domain $\epsilon_3$ between $\epsilon_{1}$ and $\epsilon_{2}$, as shown in Equation \ref{eq:interp-based-target-comp}.  

\begin{align}
    t_3 = t_1 + \frac{t_2 - t_1}{\epsilon_2 - \epsilon_1}(\epsilon_3 - \epsilon_1)
\label{eq:interp-based-target-comp}
\end{align}

where $t_i$ is the target of the current training point under $\epsilon_i$ ($i=1, 2$), $t_3$ is the approximate target under out-of-domain $\epsilon_3$, and $\epsilon_1 < \epsilon_2$.
The approximate targets are then used to find all groundtruth negative training samples under $\epsilon_3$ based on which XDT is computed.

Our evaluation shows that in most cases, Xling deploying the 
interpolation-based method has a competitive prediction quality to that using the naive solution.
Furthermore, there are two advantages of the proposed method over the naive way: (1) it is significantly faster since no range search is executed, and (2) as in Section \ref{sec:xdt-sel-eval}, interpolation-based method tends to result in no higher false negative rate (FNR) than the naive method, which is better for the effectiveness (e.g., the recall) of Xling.

\section{Experiments}
\label{sec:exps}
\subsection{Experiment settings}
\label{sec:exp-settings}
\noindent\textbf{Environment:} 
All the experiments are executed on a Lambda Quad workstation with 28 3.30GHz Intel Core i9-9940X CPUs, 4 RTX 2080 Ti GPUs and 128 GB RAM. 

\begin{table}[h]
\small
\centering
\begin{tabularx}{0.8\columnwidth}{l|l|l|l|l}
\toprule
 \textbf{Dataset} & \textbf{\#Points} & \textbf{\#Sampled}& \textbf{Dim} & \textbf{Type} \\ 
\toprule
FastText  & 1M & 150k & 300 & Text \\
Glove   & 	1.2M & 150k & 200 & 	Text\\
Word2vec   & 3M	& 150k & 300 & Text\\
Gist & 	1M & 150k & 960 &	Image\\
Sift & 	1M & 150k & 128 &	Image\\
NUS-WIDE & 	270k & 150k & 500 & 	Image\\
\bottomrule
\end{tabularx}
\caption{Evaluation dataset information, including the number of total points in the whole dataset (\textit{\#Points}), the number of sampled points for evaluation (\textit{\#Sampled}), data dimension (\textit{Dim}), and the raw data type (\textit{Type}).} 
\vspace{-2mm}
\label{tab:datasets}
\end{table} 

\noindent\textbf{Datasets: }
Table \ref{tab:datasets} provides an overview for our evaluation datasets, reporting their sizes, data dimensions and data types. We introduce more details here: 
\begin{enumerate}
    \item FastText:
    1M word embeddings (300-dimensional) generated by fastText model pre-trained on Wikipedia 2017, UMBC webbase corpus and statmt.org news dataset. 
    \item Glove:
    1.2M word vectors (200-dimensional) pre-trained on tweets. 
    \item Word2vec:
    3M word embeddings (300-dimensional) pre-trained on Google News dataset.
    \item Gist:
    1M GIST image descriptors (960-dimensional).
    \item Sift:
    1M SIFT image descriptors (128-dimensional).
    \item NUS-WIDE:
    270k bag-of-visual-words vectors (500-dimensional) learned on a web image dataset created by NUS’s Lab for Media Search.
\end{enumerate}
Due to computing resource limitation, we randomly sample 150k vectors from each of them for the evaluation. In the rest of this paper, any mentioned dataset name is by default meaning the 150k subset of the corresponding dataset.

We then normalize the sampled vectors to unit length because (1) this makes the distances bounded, i.e., both cosine and Euclidean distance are within [0, 2] on unit vectors, making it easier to determine $\epsilon$, and (2) some baseline methods do not support cosine distance, in which case we have to convert the cosine $\epsilon$ into equivalent Euclidean $\epsilon$ for them on unit vectors, as in our previous work~\cite{laf}. 
We then split each dataset into training and testing sets by a ratio of 8:2, where the training set acts as $R$ while the testing set is $S$. Xling is trained on the training set, then all the methods
are evaluated using the corresponding testing set as queries.

\noindent\textbf{Learning models:} we use several learning models to evaluate the performance of the optimization strategies in Xling. They are introduced as follows. For the deep models, we use the recommended configurations in prior works, while for non-deep models a grid search is executed to find the best parameters. Specifically, (1) RMI~\cite{case-for-learned-index-structures-kraska2018case}: We use the same configuration in \cite{laf}, i.e., three stages, respectively including 1, 2, 4 fully-connected neural networks. Each neural network has 4 hidden layers with width 512, 512, 256, and 128. The RMI is trained for 200 epochs with batch size 512. (2) NN: We also evaluate the neural network (NN), which is simply a single sub-model extracted from the RMI above. So all the parameters (including the training configuration) are same as that in RMI.
(3) SelNet~\cite{wang2021consistent}: We use the the same model configuration as in the paper. (4) XGBoost Regressor (XGB),  LightGBM Regressor (LGBM) and Support Vector Regressor (SVR): These are all non-deep regressors, we do a grid search to determine their best parameters.

\noindent\textbf{Similarity join baselines:}
The evaluation baselines include both exact and approximate methods, where the latter are the major baselines.
\begin{enumerate}
    \item Naive: This is a brute-force based nested-loop similarity join, i.e., for each query point in $S$, do a brute-force range search for it in $R$. Its results act as the groundtruth for measuring the result quality of all other methods.  
    \item SuperEGO\protect\footnote{code available at \protect\url{https://www.ics.uci.edu/~dvk/code/SuperEGO.html}} \cite{super-ego}: This is an exact method based on Epsilon Grid Ordering (EGO) that sorts the data by a specific order to facilitate the join. 
    \item LSH: This is an approximate method using LSH. First the points in $R$ are mapped into buckets by LSH, then each query is mapped to some buckets by the same LSH. The points in those buckets and nearby buckets will then be retrieved as candidates and verified. This method is implemented using FALCONN \cite{falconn}, the state-of-the-art LSH library. 
    \item KmeansTree: This is an approximate method using K-means tree, where the tree is built on $R$ and the space is partitioned and represented by sub-trees. Then each query is passed into the tree and a specific sub-tree (which represents a sub-space of $R$) will be inspected to find neighbors within the range. We use FLANN \cite{flann}, a widely used high-performance library for tree-based search, to implement this method.    
    \item Naive-LSBF: This is an approximate filter-based method that simply applies LSBF \protect\footnote{code available at \protect\url{https://github.com/csunyy/LSBF}} onto the Naive method, in the same way as the use of Xling in similarity join (Section \ref{sec:use-in-simjoin}).  We use LSBF instead of the following MSBF variants because those variants raise unnecessary overhead to support specific extra features.  
    \item IVFPQ: we also select an approximate nearest neighbor (ANN) index as part of the baselines, which is the IVFPQ index \cite{ivfadc} in FAISS \cite{faiss}, one of the industrial ANN search libraries. Since IVFPQ does not support range search natively, we evaluate it by first searching for a larger number of nearest candidates, then verifying which candidates are the true neighbors given $\epsilon$. 
    And as IVFPQ tends to achieve a high search speed with relatively lower quality (as discussed in our previous work \cite{lider}), this baseline does not make much sense in the fixed-parameter end-to-end evaluation (Section~\ref{sec:exp-end2end}). Therefore, we only evaluate it in the trade-off (Section~\ref{sec:exp-tradeoff}) and generalization (Section~\ref{sec:exp-generalization}) experiments.    
\end{enumerate}

\noindent\textbf{Our proposed methods:}
Following the way described in Section \ref{sec:use-in-simjoin}, we apply Xling to several base similarity join methods mentioned above. The resulting methods are named as: (1) XJoin (2) LSH-Xling (3) KmeansTree-Xling (4) IVFPQ-Xling. The \textit{XJoin} is our major proposed method that is evaluated in all the similarity join experiments, while the other proposed methods here are only compared with the corresponding baseline methods to show the enhancement brought by Xling to them. The learned cardinality estimator used by Xling in all these methods is an RMI. Note that the goal of this paper is to reveal the potential of such a new framework on speeding up similarity join generically, so selecting the best estimation model is out of scope. Given that RMI has been used as a strong baseline for learned cardinality estimation in \cite{mono-card-est-cardnet} and it is not the most state-of-the-art, it is a fair choice for Xling in the evaluation, especially when we can show that RMI is already good enough to outperform the other baselines. We will evaluate Xling with other estimators in the future work.

In XJoin, the XDT is computed by FPR-based selection (introduced in Section \ref{sec:xdt-sel}) with 5\% FPR tolerance, and $\tau = 50$, while in LSH-Xling, KmeansTree-Xling and IVFPQ-Xling, XDT is computed by mean-based selection with $\tau = 0$. As discussed in Section \ref{sec:xdt-sel}, ``mean-based XDT selection + smaller $\tau$'' leads to higher quality while ``FPR-based selection + larger $\tau$'' results in more acceleration. Since LSH, KmeansTree and IVFPQ are approximate methods that sacrifice quality for efficiency, mean-based XDT with lowest $\tau$ can accelerate them while minimizing the further quality loss. For Naive, the bottleneck is the efficiency, so we choose the other configuration for Xling in order to achieve a non-trivial speedup. 

\noindent\textbf{Metrics: }
The distance metric for text data is cosine distance while that for image data is Euclidean distance. For the baselines which do not support cosine distance, we follow~\cite{laf} to equivalently convert cosine distance to Euclidean distance.
The evaluation metrics include (1) end-to-end join time for measuring the efficiency of similarity join methods, (2) recall (i.e., the ratio of the returned positive results over
all the groundtruth positive results) for measuring the similarity join result quality, (3) mean absolute error (MAE) and mean squared error (MSE) for measuring prediction quality of the learned cardinality estimator, and (4) false positive rate (FPR) and false negative rate (FNR) for measuring prediction quality of Xling.

\noindent\textbf{Evaluation $\epsilon$: }
As we have normalized all the vectors, the distances between them are bounded, i.e., [0, 2] for both cosine and Euclidean distances. Since many use cases of similarity join usually choose $\epsilon$ from the range [0.2, 0.5] \cite{table-discovery-by-embedding-cosine-similarity-0.7, table-discovery-by-embedding-cosine-similarity-0.8, near-duplicate-detect-app-3-video-sim-join-with-vector, recommendation-by-embedding-cosine-similarity-0.5}, we do a grid search in this range and determine the representative evaluation $\epsilon$ values: 0.4, 0.45 and 0.5, based on the portion of negative queries (i.e., the queries having no neighbor in $R$) in $S$. We set the upper limit for the portion as 90\% as too many negative queries will make the evaluation unconvincing.
Under the selected $\epsilon$ values, most of the datasets (except NUS-WIDE) have a proper portion that is no more than 90\%.  
Table \ref{tab:portion-neg-queries} reports the portion of negative queries for each dataset under each $\epsilon$. 

\begin{table}
\small
\centering
\begin{tabular}{l|ccc}
\toprule
  \textbf{Dataset} & {\shortstack{\textbf{Portion} \\ ($\epsilon=0.4$)}} & {\shortstack{\textbf{Portion} \\ ($\epsilon=0.45$)}} & {\shortstack{\textbf{Portion} \\ ($\epsilon=0.5$)}} \\
\toprule

FastText & 0.1103 & 0.0443 & 0.0116 \\
\midrule
Glove & 0.8668 & 0.7851 & 0.6637\\
\midrule
Word2vec & 0.2875 & 0.1675 & 0.0803\\
\midrule
Gist & 0.8442 & 0.3939  & 0.1027\\
\midrule
Sift & 0.5578 & 0.3494 & 0.1531\\
\midrule
NUS-WIDE & 0.9743 & 0.9653 & 0.9544\\

\bottomrule
\end{tabular}
\caption{The portion of negative queries for each dataset under each $\epsilon$}
\label{tab:portion-neg-queries}
\vspace{-8mm}
\end{table}

\subsection{Optimization strategy evaluation}
\label{sec:optim-strat-eval}
In this section we report and analyze the evaluation results of the optimization strategies.

\subsubsection{Training $\epsilon$ selection}

\begin{table*}[h]
    %% Sift ====================================================
    \begin{subtable}[h]{0.45\textwidth}
    \tiny
        \centering
          \begin{tabular}{p{0.8cm} |p{0.8cm}|p{0.8cm}|p{0.8cm}|p{1.2cm}|p{1.3cm}}
    \hline 
    \multicolumn{2}{c|}{\textbf{Random Testing $\epsilon$}} & \multicolumn{2}{c|}{\textbf{Uniform Testing $\epsilon$}} & \multirow{2}{*}{\textbf{Model}} & \multirow{2}{*}{\textbf{Strategy}} \\ \cline{1-4}
      \textbf{MAE}  & \textbf{MSE} & \textbf{MAE}  & \textbf{MSE} &  &\\ \hline
    4.04 & 6.69  & 2.57 & 2.68 & XGB & fixed\\
    \textbf{2.54} & \textbf{1.32} & \textbf{2.54} & \textbf{1.26} &  XGB &  \textbf{auto}\\
    \midrule
    3.66 & 6.65 &  2.14 & 2.57 &  LGBM & fixed\\
    \textbf{1.8} & \textbf{0.83} & \textbf{1.79} & \textbf{0.81} &  LGBM & \textbf{auto}\\
    \midrule
    23.79 & 97.16 & 27.55 & 127.15 & SVR  & fixed\\
    \textbf{22.26} & \textbf{81.06} & \textbf{25.34} & \textbf{106.22} &  SVR  & \textbf{auto}\\
    \midrule
    96.13 & 24711.7  & 96.44 & 28299.85 & SelNet  & fixed\\
    \textbf{93.8} & \textbf{22228.79} & \textbf{94.39} & \textbf{25737.67} & SelNet  & \textbf{auto}\\
    \midrule
    2.08 & 1.39 & 1.69 & 1.00 & NN & fixed\\
    \textbf{0.39} & \textbf{0.03} & \textbf{0.44} & \textbf{0.05} &  NN &  \textbf{auto}\\ 
    \midrule
    2.55 & 5.72 &  1.52 & 2.18 & RMI  & fixed\\
    \textbf{0.20} & \textbf{0.01} & \textbf{0.19} & \textbf{0.01} &  RMI  & \textbf{auto}\\
    \hline
    \end{tabular}
        \caption{Sift}
        \label{tab:sub_exp1_Sift}
    \end{subtable}
    \hfill
    %% Word2Vec ====================================================
    \begin{subtable}[h]{0.45\textwidth}
    \tiny
        \centering
          \begin{tabular}{p{0.8cm} |p{0.8cm}|p{0.8cm}|p{0.8cm}|p{1.2cm}|p{1.3cm}}
    \hline 
    \multicolumn{2}{c|}{\textbf{Random Testing $\epsilon$}} & \multicolumn{2}{c|}{\textbf{Uniform Testing $\epsilon$}} & \multirow{2}{*}{\textbf{Model}} & \multirow{2}{*}{\textbf{Strategy}} \\ \cline{1-4}
      \textbf{MAE}  & \textbf{MSE} & \textbf{MAE}  & \textbf{MSE} &  &\\ \hline
    5.32 & \textbf{22.84} & \textbf{7.81} & \textbf{48.16} & XGB  & \textbf{fixed}\\
    \textbf{5.02} & 24.09 & 7.89 & 50.7 & XGB  & auto\\
    \midrule
    5.29 & 20.46 & 7.45 & 43.82 & LGBM  & fixed\\
    \textbf{4.49} & \textbf{19.61}  & \textbf{7.1} & \textbf{43.52} & LGBM & \textbf{auto}\\
    \midrule
    9.01 & \textbf{46.41} & 12.6 & \textbf{84.58} & SVR  & fixed\\
    \textbf{8.7} & 50.18 & \textbf{12.32} & 90.46 & SVR& \textbf{auto}\\
    \midrule
    \textbf{8.91} & \textbf{54.69}  & \textbf{12.57} & \textbf{97.23} & SelNet  & \textbf{fixed}\\
    8.91 & 54.7 & 12.57 & 97.24 & SelNet  & auto\\
    \midrule
    4.38 & 18.87 & 6.78 & 41.95 & NN & fixed\\
    \textbf{3.33} & \textbf{11.85} & \textbf{5.09} & \textbf{23.19} &  NN  & \textbf{auto}\\
    \midrule
    4.52 & 19.44 & 6.85 & 42.57 & RMI  & fixed\\
    \textbf{3.43} & \textbf{13.84} & \textbf{5.35} & \textbf{27.51} &  RMI  & \textbf{auto}\\
    \hline
    \end{tabular}
        \caption{Word2Vec}
        \label{tab:sub_exp1_Word2vec}
    \end{subtable}
    
    %% Fasttext ====================================================
    \begin{subtable}[h]{0.45\textwidth}
    \tiny
        \centering
          \begin{tabular}{p{0.8cm}|p{0.8cm}|p{0.8cm}|p{0.8cm}|p{1.2cm}|p{1.3cm}}
    \hline 
    \multicolumn{2}{c|}{\textbf{Random Testing $\epsilon$}} & \multicolumn{2}{c|}{\textbf{Uniform Testing $\epsilon$}} & \multirow{2}{*}{\textbf{Model}} & \multirow{2}{*}{\textbf{Strategy}} \\ \cline{1-4}
      \textbf{MAE}  & \textbf{MSE} & \textbf{MAE}  & \textbf{MSE} &  &\\ \hline
    4.69 & 5.49 & 5.02 & 5.36 & XGB & fixed\\
    \textbf{2.53} & \textbf{1.43} & \textbf{2.55} & \textbf{1.49} &  XGB &  \textbf{auto}\\
    \midrule
    4.7 & 5.49 &  4.76 & 4.94  & LGBM  & fixed\\
    \textbf{2.45} & \textbf{1.28}  & \textbf{2.47} & \textbf{1.33} &  LGBM  & \textbf{auto}\\
    \midrule
    27.61 & 141.55 & 29.89 & 169.73 & SVR  & fixed\\
    \textbf{23.45} & \textbf{81.06}  & \textbf{24.5} & \textbf{94.06} &  SVR  & \textbf{auto}\\
    \midrule
    36.19 & 357.57 & 38.14 & 387.45 & SelNet  & fixed\\
    \textbf{36.19} & \textbf{357.52}  & \textbf{38.14} & \textbf{387.39} & SelNet  & \textbf{auto}\\
    \midrule
    0.54 & 0.14 & 0.79 & 0.36 & NN & fixed\\
    \textbf{0.32} & \textbf{0.03} & \textbf{0.38} & \textbf{0.05} &  NN  & \textbf{auto}\\
    \midrule
    2.09 & 2.23 & 2.80 & 4.49 & RMI  & fixed\\
    \textbf{0.67} & \textbf{0.18} & \textbf{0.77} & \textbf{0.24} &  RMI  & \textbf{auto}\\
    \hline
    \end{tabular}
        \caption{FastText}
        \label{tab:sub_exp1_Fasttext}
    \end{subtable}
    \hfill
    %% NUS-WIDE ====================================================
    \begin{subtable}[h]{0.45\textwidth}
    \tiny
        \centering
          \begin{tabular}{p{0.8cm} |p{0.8cm}|p{0.8cm}|p{0.8cm}|p{1.2cm}|p{1.3cm}}
    \hline 
    \multicolumn{2}{c|}{\textbf{Random Testing $\epsilon$}} & \multicolumn{2}{c|}{\textbf{Uniform Testing $\epsilon$}} & \multirow{2}{*}{\textbf{Model}} & \multirow{2}{*}{\textbf{Strategy}} \\ \cline{1-4}
      \textbf{MAE}  & \textbf{MSE} & \textbf{MAE}  & \textbf{MSE} &  &\\ \hline
    4.74 & 14.99 &  3.4 & 6.35 &  XGB  & fixed\\
    \textbf{3.36} & \textbf{2.62}& \textbf{3.37} & \textbf{2.69} & XGB &  \textbf{auto}\\
    \midrule
    4.23 & 15.29 & 2.89 & 6.4 & LGBM  & fixed\\
    \textbf{2.4} & \textbf{2.6} & \textbf{2.36} & \textbf{2.46} &  LGBM &  \textbf{auto}\\
    \midrule
    \textbf{22.81} & \textbf{70.71}  & \textbf{24.6} & \textbf{80.33} &  SVR  & \textbf{fixed}\\
    25.22 & 99.4  & 25.06 & 97.04  & SVR  & auto\\
    \midrule
    82.26 & 8655.88  & 82.62 & 9886.03  & SelNet  & fixed\\
    \textbf{75.96} & \textbf{6560.3} & \textbf{75.93} & \textbf{7490.16} & SelNet  & \textbf{auto}\\
    \midrule
    3.39 & 10.56 & 2.51 & 5.01 & NN & fixed\\
    \textbf{0.83} & \textbf{0.20} & \textbf{1.04} & \textbf{0.33} &  NN  & \textbf{auto}\\
    \midrule
    4.24 & 19.66 & 2.49 & 6.60 & RMI  & fixed\\
    \textbf{0.48} & \textbf{0.27} & \textbf{0.39} & \textbf{0.13} &  RMI  & \textbf{auto}\\
    \hline
    \end{tabular}
        \caption{NUS-WIDE}
        \label{tab:sub_exp1_Nus_wide_bow}
    \end{subtable}
    \caption{Effectiveness of the two training $\epsilon$ selection strategies for different regressors trained on different datasets, where \textit{MAE} columns show the original numbers multiplied by $10^{-3}$ and \textit{MSE} columns show the original numbers multiplied by  $10^{-7}$ }
    \label{tab:sub_exp1}
\end{table*}

The results of the training $\epsilon$ selection strategy is reported in Table \ref{tab:sub_exp1}. Due to the space limit, we only present the results for 4 out of 6 datasets.

Following Definition~\ref{def:train-cond-sel}, to make it simple, we have the set of candidate condition values shared by all the training points, i.e., the set \{$c_{i1}, c_{i2}, ..., c_{im}$\} is same for any training point $p_i$. We let $m = 100$ and construct such a set by evenly sampling 100 values from a range [$c_{min}, c_{max}$], where we set $c_{min} = 0.4, c_{max} = 0.9$ for cosine distance while $c_{min} = 0.5, c_{max} = 2.0$ for Euclidean distance. These $c_{min}$ and $c_{max}$ values are selected based on a grid search given the evaluation experience from our previous work \cite{laf}. Then we set the sampling number $s = 6$, i.e., for each training point, 6 distinct condition values will be sampled from the 100 candidate values and become the final training $\epsilon$ values to be paired with the point in the prepared training data. Two sampling strategies are deployed to select the 6 values: (1) uniform sampling, e.g., selecting the 1st, 20th, 40th, 60th, 80th and 100th values (2) our ATCS strategy (Algorithm\ref{alg:train-cond-sel}). The former is presented as ``fixed'' while the latter is marked as ``auto'' in the \textit{Strategy} columns of Table~\ref{tab:sub_exp1}.

The regression models (i.e., the learning models listed in Section~\ref{sec:exp-settings}) will first be trained respectively using the two prepared training sets from different strategies, then they will do inference on some prepared testing sets and the inference quality is measured by MAE and MSE, which can reflect the training effectiveness. For a fair testing, we generate two kinds of prepared testing sets: (1) the testing points with randomly selected $\epsilon$ values from the 100 candidates mentioned above, and (2) the same testing points with uniformly selected $\epsilon$ values from the candidates. The inference quality on the former set is reported in the \textit{Random Testing $\epsilon$} columns while that on the latter is reported in the \textit{Uniform Testing $\epsilon$} columns of Table~\ref{tab:sub_exp1}. And for each learning model, the \textit{Strategy} which results in better inference quality (i.e., lower MAE and MSE) is highlighted by bold text.     

In most cases, the ``auto'' strategy (i.e., our ATCS strategy) achieves a better inference quality (e.g., reducing up to 98\% of the MSE on NUS-WIDE dataset) than uniform training $\epsilon$,
meaning that ATCS strategy is generic to facilitate various kinds of regression models and highly effective to raise a significant improvement on the prediction quality. Therefore, for all the following experiments in this paper, the training sets are prepared by ATCS.

\subsubsection{XDT selection}
\label{sec:xdt-sel-eval}

\begin{table*}[h]
\tiny
    \centering
      \begin{tabular}{l|c|c|c|c|c|c|c|c|c|c|c}
\hline
\multirow{2}{*}{\textbf{Dataset}} & \multirow{2}{*}{\textbf{Model}} &  \multirow{2}{*}{\textbf{$\epsilon$}} & \multirow{2}{*}{\textbf{XDT Selection}}  & \multicolumn{4}{c|}{ \textbf{Approximate Targets}} & \multicolumn{4}{c}{\textbf{Exact Targets}}  \\ 
\cline{5-12}
 &  & & & FPR & FNR & \textbf{Time(s)} & XDT & FPR & FNR & \textbf{Time(s)} & XDT  \\ 
% Glove
\toprule
\multirow{12}{*}{Glove} 
& \multirow{6}{*}{XGB} 
& \multirow{2}{*}{0.4} 
& mean & 0.4948 & 0.3982 & 1.3618 & -8.45 & 0.4948 & 0.3982 & 1037.9202 & -8.45\\
& & & FPR & 0.0562 & 0.8578 & 1.3859 & 406.83 & 0.0562 & 0.8578 & 1037.9278 & 406.83 \\
\cline{3-12}
& & \multirow{2}{*}{0.45} 
& mean & 0.5019 & 0.4247 & 2.5836 & -10.66 & 0.4996 & 0.4259 & 1038.6554 &  -9.28   \\
& & & FPR & 0.0608 & 0.8737 & 2.5499 &  392.43 & 0.0548 & 0.882 & 1038.6161 & 405.70 \\
\cline{3-12}
& & \multirow{2}{*}{0.5} 
&  mean & 0.4989 & 0.4373 & 2.51 & -12.56 & 0.5004 & 0.4358 & 1038.3489 & -13.50\\
& & & FPR & 0.0602 & 0.8903 & 2.8515 &  387.51 & 0.0566 & 0.8946 & 1038.379 &  396.47 \\
\cline{2-12}
& \multirow{6}{*}{RMI} 
& \multirow{2}{*}{0.4} 
&  mean  & 0.4071 & 0.2506 & 10.0679 & 4.38 & 0.4071 & 0.2506 & 1046.1885 & 4.38\\
& & & FPR  & 0.0501 & 0.5049 & 9.9112 & 6.46 & 0.0501 & 0.5049 & 1046.1892  & 6.46\\
\cline{3-12}
& & \multirow{2}{*}{0.45} 
& mean & 0.4496 & 0.2308 & 10.9286 & 4.57 & 0.3537 & 0.2702 & 1046.9525  & 4.76  \\
& & & FPR  & 0.0808 & 0.4902 & 11.1269 & 6.48 & 0.0506 & 0.5385 & 1047.0098 & 7.37 \\
\cline{3-12}
& & \multirow{2}{*}{0.5} 
& mean  & 0.3613 & 0.2546 & 10.8683 & 5.15 & 0.3043 & 0.284 & 1046.6399   & 5.34\\
& & & FPR & 0.0709 & 0.5041 & 11.1455 & 7.99 & 0.0518 & 0.5424 & 1046.5916 & 8.82 \\

\cline{1-12}

% NUS-WIDE
\multirow{12}{*}{NUS-WIDE} 
& \multirow{6}{*}{XGB} 
& \multirow{2}{*}{0.4} 
& mean & 0.5041 & 0.2202 & 1.4501 & -3.31 & 0.5011 & 0.2215 & 2638.2846 & 15.41\\
& & & FPR & 0.0555 & 0.6554 & 1.5899 &  3424.20 & 0.0524 & 0.6645 & 2638.2916 & 3499.68 \\
\cline{3-12}
& & \multirow{2}{*}{0.45} 
& mean & 0.5024 & 0.2428 & 1.4116 &  -3.31 & 0.5008 & 0.2428 & 2655.1167 & 5.55  \\
& & & FPR & 0.0535 & 0.6727 & 1.4341 &  3424.20 & 0.0519 & 0.6756 & 2655.1188  & 3463.20\\
\cline{3-12}
& & \multirow{2}{*}{0.5} 
&  mean & 0.5009 & 0.2719 & 1.8672 & -3.31 & 0.5009 & 0.2719 & 2651.2069 & -3.31 \\
& & & FPR & 0.0516 & 0.6981 & 2.0873 & 3424.20 & 0.0516 & 0.6981 & 2651.1754 & 3424.20\\
\cline{2-12}
& \multirow{6}{*}{RMI} 
& \multirow{2}{*}{0.4} 
&  mean  & 0.3672 & 0.057 & 10.6236 & 2.86 & 0.3537 & 0.0596 & 2647.5025 & 2.99\\
& & & FPR  & 0.0534 & 0.0894 & 11.0576 & 9.87 & 0.048 & 0.092 & 2647.5929 & 10.45 \\
\cline{3-12}
& & \multirow{2}{*}{0.45} 
& mean & 0.3301 & 0.071 & 10.4377 & 2.43 & 0.3216 & 0.071 & 2664.379 & 2.52 \\
& & & FPR  & 0.0504 & 0.1123 & 10.8119 & 9.55 & 0.0463 & 0.1161 & 2664.2925 & 9.95 \\
\cline{3-12}
& & \multirow{2}{*}{0.5} 
& mean  & 0.2872 & 0.087 & 11.3239 & 2.18 & 0.2872 & 0.087 & 2660.6731 & 2.18\\
& & & FPR & 0.0467 & 0.1294 & 11.3493 & 9.54 & 0.0467 & 0.1294 & 2660.3748 & 9.54 \\
\bottomrule
\end{tabular}
    \caption{Prediction quality (FPR and FNR) of Xling when XDT is computed in different ways (while fixing $\tau = 0$). There are two dimensions for a way to compute XDT: (1) using mean-based or FPR-based XDT selection method (2) using interpolation-based method to approximate the targets or naive method to compute the exact targets. The time to compute the targets is also presented (\textit{Time}). Results on other datasets are similar to Glove and NUS-WIDE. Note that we allow XDT to be less than zero.}
    \label{tab:sub_exp2_2}
\end{table*}

As mentioned in Section \ref{sec:xdt-sel}, XDT is influenced by three factors: (1) $\tau$, (2) the XDT selection method 
 (mean-based or FPR-based), and (3) the way to get training targets for out-of-domain $\epsilon$ (interpolation-based approximate targets or naively computed exact targets). In this section we evaluate the last two factors due to page limit.
 Specifically, here we fix $\tau = 0$. For each dataset, the learned cardinality estimator of Xling is first trained using the training set, then we vary the setting for the two factors, compute XDT under each setting, and measure the FPR and FNR of Xling on the testing set of the current dataset, as well as the time for computing the training targets.

Due to space limit, Table~\ref{tab:sub_exp2_2} reports the results for two representative models (i.e., XGB for non-deep and RMI for deep model) on two datasets Glove (text) and NUS-WIDE (image). For the factor of target computing, the results show that the two target computing methods usually lead to similar FPR and FNR, while our proposed interpolation-based target approximation is around 100x $\sim$ 2000x faster than the naive target computing, meaning that interpolation-based target approximation is both effective and highly efficient. For the factor of XDT selection, as mentioned in Section~\ref{sec:xdt-sel}, the results present that FPR-based XDT selection usually generates a higher XDT than mean-based, thereby the former tends to determine more queries as negative and causes lower FPR and higher FNR than the latter. So in the end-to-end similarity join, using FPR-based XDT selection will lead to higher speedup but lower quality than mean-based, which supports the statements in Section~\ref{sec:xdt-sel}.    

\subsection{Filter effectiveness evaluation}
In this section we evaluate the prediction quality between Xling and LSBF, reported by FPR and FNR of their predictions for the testing sets. We fix $\tau = 0$ as LSBF does not support the cases where $\tau > 0$.

The results are included in Table~\ref{tab:quality-lsbf-vs-xling}. In addition to FPR and FNR, the table also includes the total number of true neighbors found for all the testing queries (\textit{\#Nbrs}) by Naive-LSBF and XJoin, the number of predicted positive queries (\textit{\#PPQ}) which are the query
points predicted as positive by the filter, and the average number of neighbors per predicted positive query (\textit{\#ANPQ}), i.e., \#ANPQ = \#Nbrs/\#PPQ. We have the following observations: (1) In most cases, Xling with mean-based XDT has both lower FPR and FNR than LSBF, meaning it is more effective than LSBF.
(2) In many cases, Xling with FPR-based XDT performs similarly to the observation above, while in some other cases, it has higher FNR than LSBF. However, even with a higher FNR, it still finds more true neighbors in those cases than LSBF, e.g., the case of Word2vec under $\epsilon=0.5$. The reason is that the queries predicted as positive by Xling overall have more neighbors than those by LSBF (reflected by \#ANPQ), due to which Xling can find more true results with less range search (i.e., the \#PPQ of FPR-based Xling is usually less than LSBF). In short, Xling learns the data distribution and makes predictions based on the data density, such that it can maximize the recall with minimized number of search, which is a huge advantage over the data-unaware LSBF.

\begin{table*}
    \begin{subtable}[h]{0.47\textwidth}
    \tiny
    \centering
    \begin{tabular}{p{1cm} |p{0.4cm}|p{1.3cm}|p{0.6cm}p{0.6cm}p{0.5cm}p{0.5cm}p{0.7cm}}
    \toprule
      \textbf{Dataset} & \textbf{$\epsilon$} & \textbf{Filter} & \textbf{FPR} & \textbf{FNR} & {\shortstack{\textbf{\#Nbrs} \\ ($\times10^5$)}} & {\shortstack{\textbf{\#PPQ} \\ ($\times10^4$)}} & \textbf{\#ANPQ}\\ 
    \toprule
    
    \multirow{9}{*}{FastText} 
    & \multirow{3}{*}{0.4} 
    & LSBF & 0.46 & 0.2908 & 127.73 & 2.05 & 624.58\\
    & & Xling(mean) & 0.2523 & 0.1114 & 142.15 & 2.46 & 578.95\\
    & & Xling(FPR) & 0.0529 & 0.207 & 141.90 & 2.13 & 664.94\\
    \cline{2-8}
    & \multirow{3}{*}{0.45} 
    & LSBF & 0.5124 & 0.2471 & 296.05 & 2.23 & 1329.56\\
    & & Xling(mean) & 0.3469 & 0.0682 & 325.25 & 2.72 & 1196.82\\
    & & Xling(FPR) & 0.0948 & 0.1295 & 325.07 & 2.51 & 1295.92\\
    \cline{2-8}
    & \multirow{3}{*}{0.5} 
    & LSBF & 0.5244 & 0.2062 & 708.64 & 2.37 & 2987.66\\
    & & Xling(mean) & 0.3582 & 0.0376 & 767.57 & 2.87 & 2678.19\\
    & & Xling(FPR) & 0.0802 & 0.099 & 767.23 & 2.68 & 2868.68\\
    \midrule
    
    \multirow{9}{*}{Word2vec} 
    & \multirow{3}{*}{0.4} 
    & LSBF & 0.4952 & 0.4607 & 3.49 & 1.58 & 22.06\\
    & & Xling(mean) & 0.374 & 0.3197 & 4.63 & 1.78 & 26.08\\
    & & Xling(FPR) & 0.0502 & 0.7028 & 4.20 & 0.68 & 61.81\\
    \cline{2-8}
    & \multirow{3}{*}{0.45} 
    & LSBF & 0.5516 & 0.3977 & 6.77 & 1.78 & 37.98\\
    & & Xling(mean) & 0.4397 & 0.234 & 8.86 & 2.13 & 41.53\\
    & & Xling(FPR) & 0.0804 & 0.5996 & 8.19 & 1.04 & 78.70\\
    \cline{2-8}
    & \multirow{3}{*}{0.5} 
    & LSBF & 0.5943 & 0.3419 & 13.04 & 1.96 & 66.56\\
    & & Xling(mean) & 0.3019 & 0.2725 & 16.40 & 2.08 & 78.83\\
    & & Xling(FPR) & 0.039 & 0.5882 & 15.15 & 1.15 & 132.22\\\bottomrule
    \end{tabular}
    \caption{Text datasets}
    \label{tab:quality-lsbf-vs-xling-on-text-ds}
    \end{subtable}
    \hfill
    \begin{subtable}[h]{0.47\textwidth}
    \tiny
    \centering
    \begin{tabular}{p{1.1cm} |p{0.4cm}|p{1.3cm}|p{0.5cm}p{0.5cm}p{0.5cm}p{0.5cm}p{0.7cm}}
    \toprule
      \textbf{Dataset} & \textbf{$\epsilon$} & \textbf{Filter} & \textbf{FPR} & \textbf{FNR} & {\shortstack{\textbf{\#Nbrs} \\ ($\times10^5$)}} & {\shortstack{\textbf{\#PPQ} \\ ($\times10^4$)}} & \textbf{\#ANPQ}\\ 
    \toprule

    \multirow{9}{*}{Sift} 
    & \multirow{3}{*}{0.4} 
    & LSBF & 0.2415 & 0.6609 & 4.69 & 0.85 & 54.89\\
    & & Xling(mean) & 0.06 & 0.6119 & 7.18 & 0.62 & 116.65\\
    & & Xling(FPR) & 0.0351 & 0.6771 & 7.08 & 0.49 & 145.37\\
    \cline{2-8}
    & \multirow{3}{*}{0.45} 
    & LSBF & 0.2338 & 0.6358 & 13.04 & 0.96 & 136.42\\
    & & Xling(mean) & 0.081 & 0.4565 & 19.49 & 1.15 & 170.09\\
    & & Xling(FPR) & 0.0537 & 0.5064 & 19.39 & 1.02 & 190.19\\
    \cline{2-8}
    & \multirow{3}{*}{0.5} 
    & LSBF & 0.2053 & 0.6139 & 35.54 & 1.08 & 330.55\\
    & & Xling(mean) & 0.1798 & 0.237 & 52.83 & 2.02 & 261.36\\
    & & Xling(FPR) & 0.0581 & 0.3678 & 52.50 & 1.63 & 321.55\\\midrule
    
    \multirow{9}{*}{NUS-WIDE} 
    & \multirow{3}{*}{0.4} 
    & LSBF & 0.2492 & 0.5687 & 1.39 & 0.76 & 18.19\\
    & & Xling(mean) & 0.3671 & 0.057 & 2.14 & 1.15 & 18.65\\
    & & Xling(FPR) & 0.0534 & 0.0894 & 2.14 & 0.23 & 94.32\\
    \cline{2-8}
    & \multirow{3}{*}{0.45} 
    & LSBF & 0.2498 & 0.5374 & 2.82 & 0.77 & 36.53\\
    & & Xling(mean) & 0.3302 & 0.071 & 4.06 & 1.05 & 38.56\\
    & & Xling(FPR) & 0.0503 & 0.1123 & 4.06 & 0.24 & 170.27\\
    \cline{2-8}
    & \multirow{3}{*}{0.5} 
    & LSBF & 0.2469 & 0.5029 & 5.15 & 0.78 & 66.42\\
    & & Xling(mean) & 0.2873 & 0.087 & 6.98 & 0.95 & 73.69\\
    & & Xling(FPR) & 0.0467 & 0.1294 & 6.98 & 0.25 & 276.09\\
    \bottomrule
    \end{tabular}
    \caption{Image datasets}
    \label{tab:quality-lsbf-vs-xling-on-image-ds}
    \end{subtable}
    \caption{The prediction quality of LSBF and Xling (mean-based or FPR-based) on different datasets, where \textit{\#Nbrs} is the total number of returned neighbors for all the queries, \textit{\#PPQ} stands for the number of \textit{\underline{P}redicted \underline{P}ositive \underline{Q}ueries}, i.e., the query points predicted as positive by the filter, and \textit{\#ANPQ} presents the \textit{\underline{A}verage number of \underline{N}eighbors per predicted \underline{P}ositive \underline{Q}uery} that equals \#Nbrs over \#PPQ. Due to space limit, the results on Glove and Gist datasets are hidden, which are similar to the other four. }
    \label{tab:quality-lsbf-vs-xling}
    \vspace{-6mm}
\end{table*}

\subsection{End-to-end evaluation}
\label{sec:exp-end2end}
In end-to-end evaluation, the key parameters for the methods are fixed as such: (1) Naive and SuperEGO have no user-specific parameters. (2) In LSH, the number of hash tables $l = 10$, number of has functions $k = 18$, the number of hash buckets to be inspected in each table $n_p = 40$ for text datasets while $n_p = 2$ for image datasets. (3) In KmeansTree, the branching factor is fixed to be 3, the portion of leaves to be inspected $\rho = 0.02$ for text datasets while $\rho = 0.012$ for image datasets. (4) For Naive-LSBF, its LSH-related parameters $k$ and $l$ are the same as the method LSH, length of the bit array in LSBF is fixed to be 2,160,000 (i.e., $|R|\times k$), the parameter $W$ in the p-stable LSH functions is set to be 2.5 for text datasets while $W = 2$ for image datasets. 
(5) For IVFPQ, the parameters for building the index are $C = 300$, $m = 32$ or $m=25$ when the data dimension is not an integer multiple of 32, $b = 8$, $p = 50$, where $C$ is the number of clusters, $m$ is the number of segments into which each data vector will be split, $b$ is the number of bits used to encode the clsuter centroids, and $p$ is the number of nearest clusters to be inspected during search. In our implementation, this baseline first uses the IVFPQ index to select 1000 nearest neighbors then verifies them given $\epsilon$. The number 1000 is determined since for all datasets, at least 50\% testing queries have less than 1000 neighbors within $\epsilon$, and in most datasets 1000 candidates are enough to find all correct neighbors for 70\% $\sim$ 90\% testing queries.
(6) The configuration of Xling is introduced in Section~\ref{sec:exp-settings}, while the base methods in SuperEGO-Xling, LSH-Xling and KmeansTree-Xling use the same parameters as above.

\begin{figure*}[h]%
    \centering
    \subfloat{{
        \label{fig:legend_chart_e2e}
        \includegraphics[width=1\textwidth]{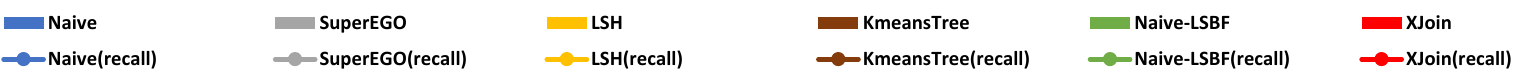}
    }}%
    \qquad
    \subfloat[\centering FastText]{{
        \label{fig:exp_end2end_fasttext}
        \includegraphics[width=0.25\textwidth]{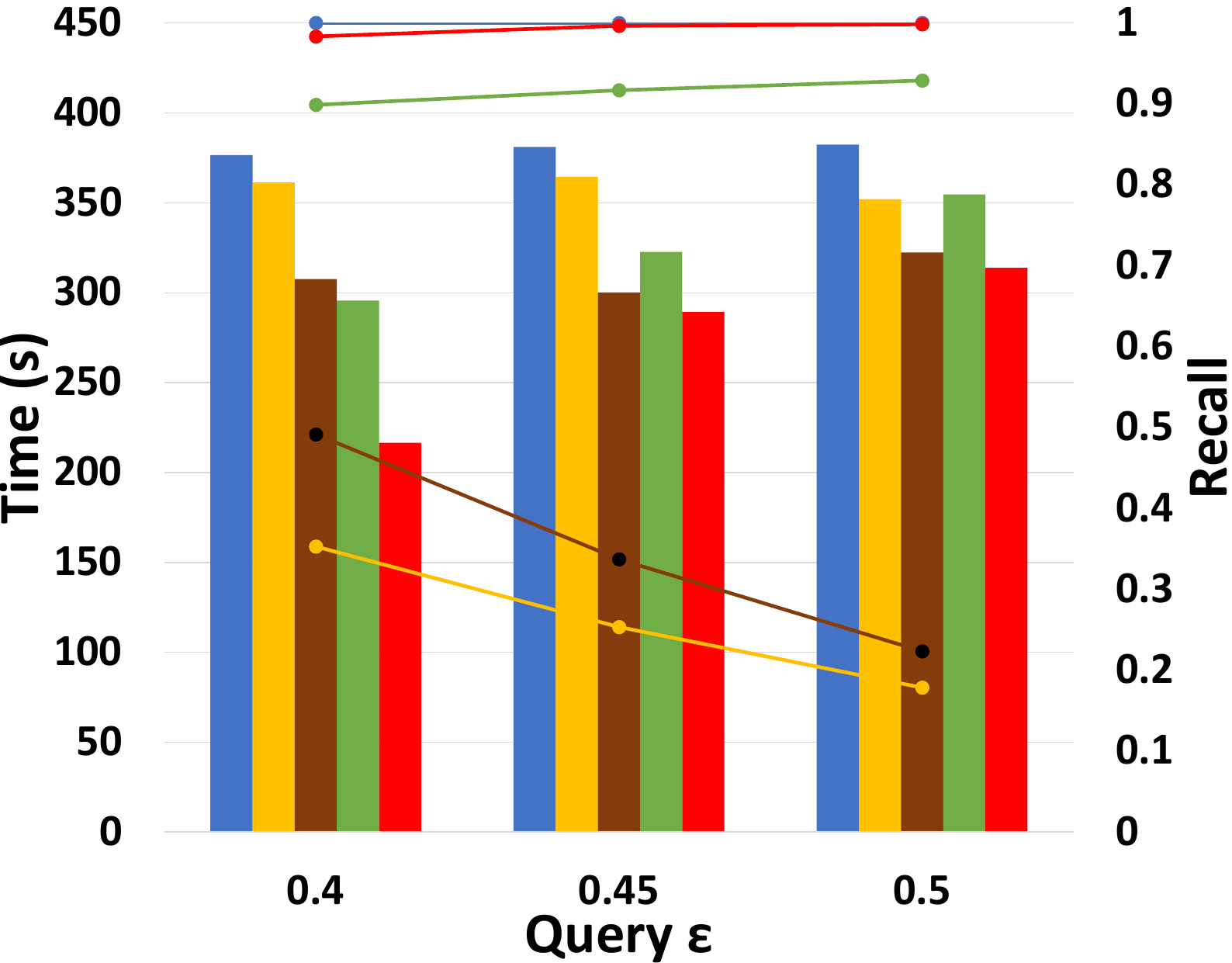}
    }}%
    \qquad
    \subfloat[\centering Glove]{{
        \label{fig:exp_end2end_glove}
       \includegraphics[width=0.25\textwidth]{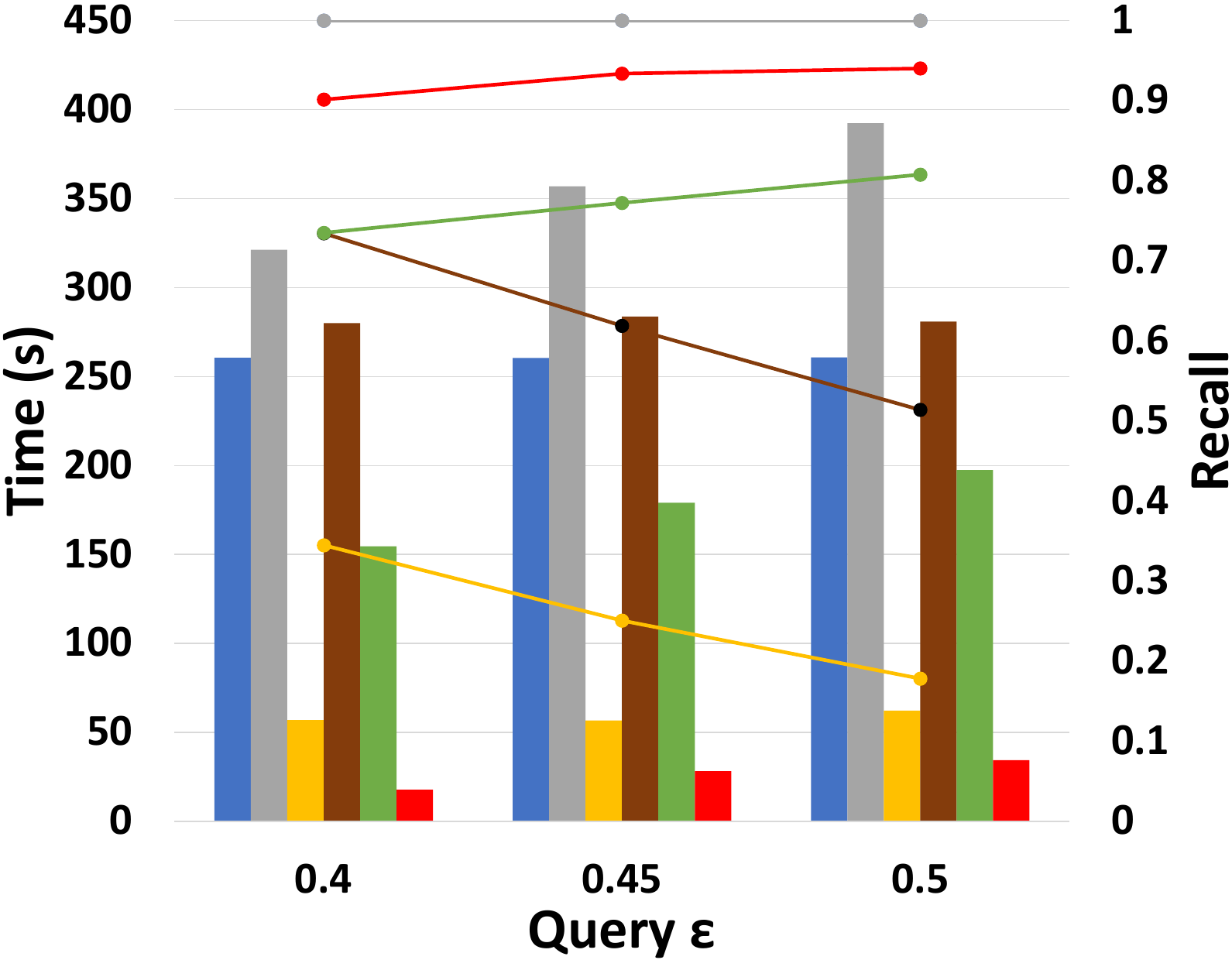}
    }}%
    \qquad
    \subfloat[\centering Word2vec]{{
        \label{fig:exp_end2end_word2vec}
        \includegraphics[width=0.25\textwidth]{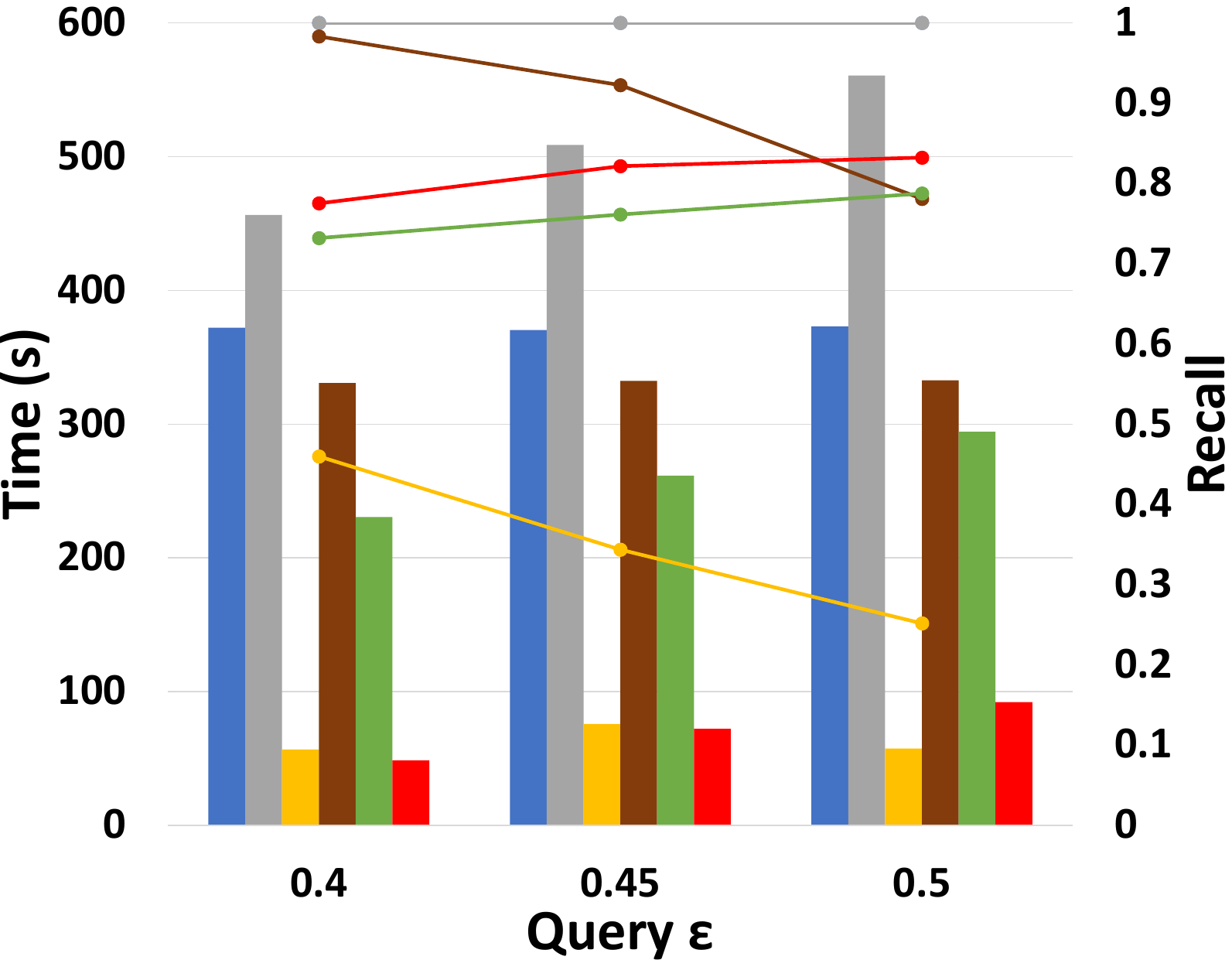}
    }}%
    \qquad
    \subfloat[\centering Gist]{{
        \label{fig:exp_end2end_gist}
        \includegraphics[width=0.25\textwidth]{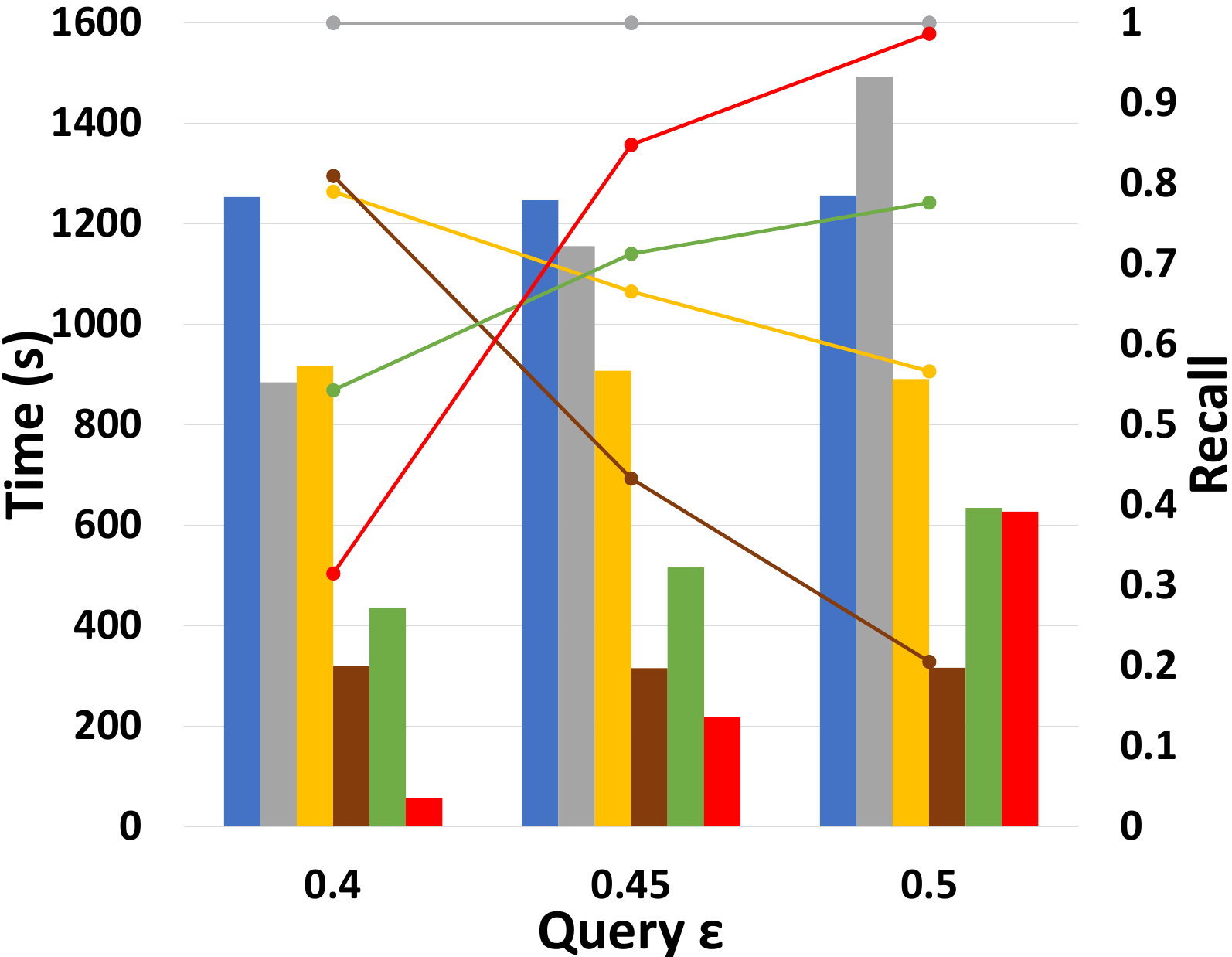}
    }}%
    \qquad
    \subfloat[\centering Sift]{{
        \label{fig:exp_end2end_sift}
        \includegraphics[width=0.25\textwidth]{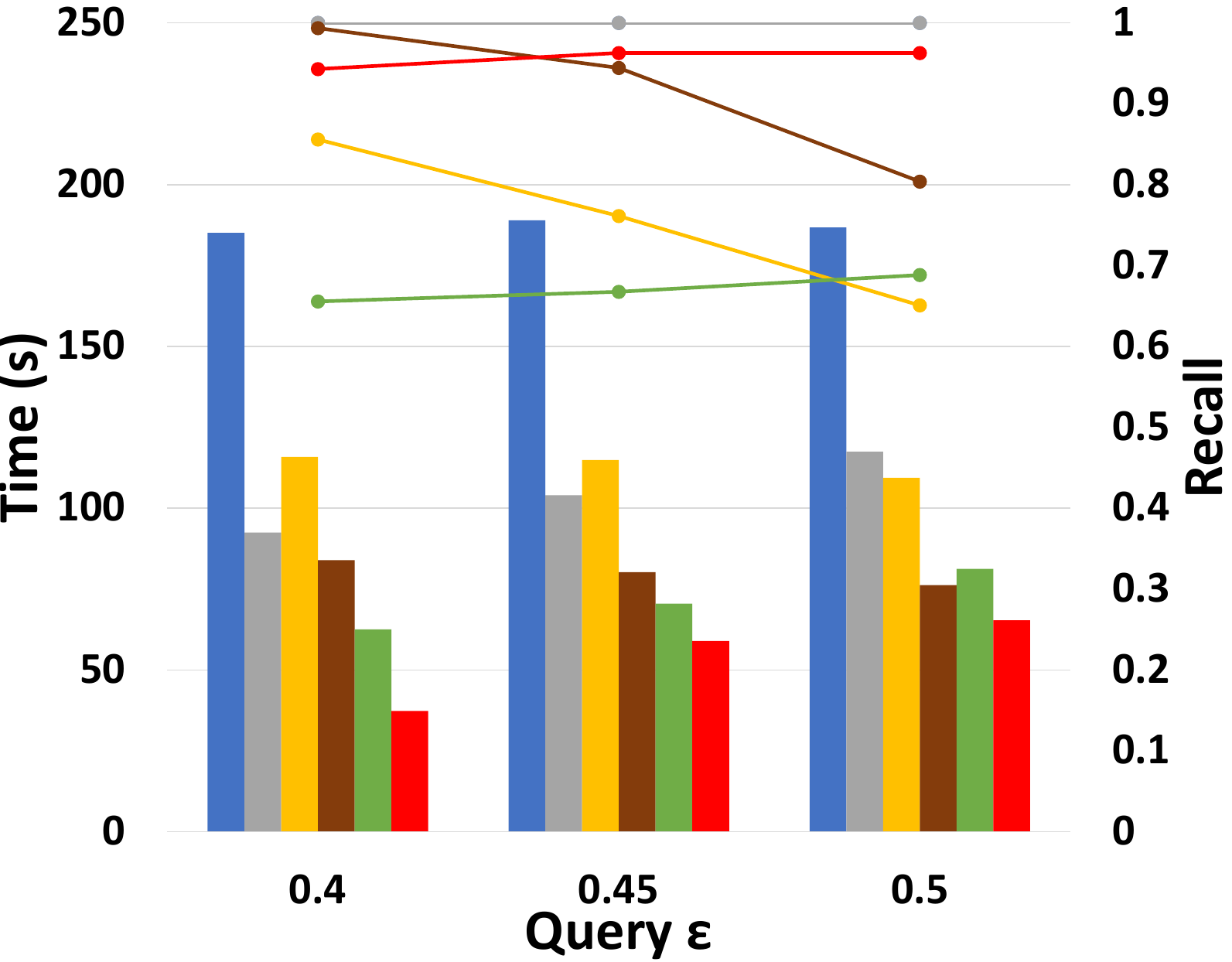}
    }}%
    \qquad
    \subfloat[\centering NUS-WIDE]{{
        \label{fig:exp_end2end_nus_wide}
       \includegraphics[width=0.25\textwidth]{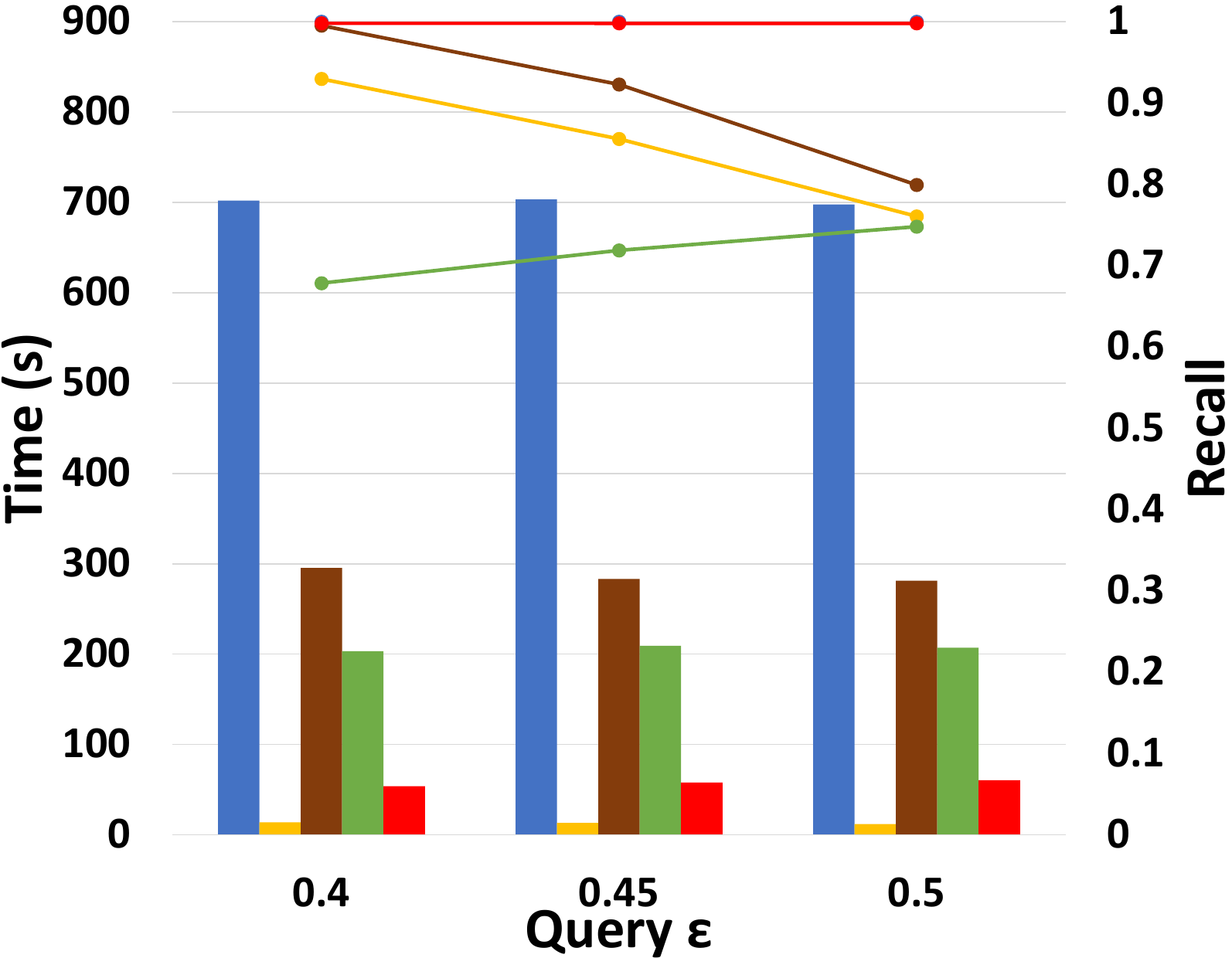}
    }}
    \caption{End-to-end query processing time and recall for all the similarity join methods on all datasets, where the figures of FastText and NUS-WIDE do not include SuperEGO, as it cannot run on these two datasets.}%
    \label{fig:end2end-exps}%
\end{figure*}

The results are illustrated in Figure~\ref{fig:end2end-exps}. SuperEGO has unknown bugs that prevent it from running on FastText and NUS-WIDE datasets, so Figure~\ref{fig:end2end-exps} does not include SuperEGO for these two datasets. SuperEGO relies on multi-threading for acceleration, so on some datasets its running time is even longer than the Naive method since we only use one thread to evaluate all the methods. The results show that our method XJoin has a higher speed (presented by the red bar) than all the baseline methods, as well as higher quality (reported by the red line) than the other approximate methods (i.e., LSH,  KmeansTree and Naive-LSBF)  in most cases. The recall of Naive and SuperEGO is always 1 as they are exact methods. Specifically, XJoin achieves up to 17x and 14.6x speedup respectively compared to the exact methods and approximate methods while guaranteeing a high quality. The results prove the high efficiency and effectiveness of our proposed method.

\subsection{Speed-quality trade-off evaluation}
\label{sec:exp-tradeoff}

\begin{figure*}[h]%
    \centering
    \subfloat{{
        \label{fig:legend_tradeoff}
        \includegraphics[width=1\textwidth]{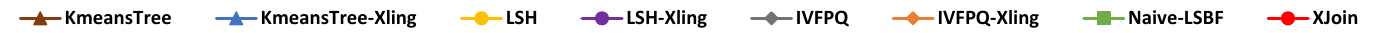}
    }}%
    \qquad
    \subfloat[\centering Glove]{{
        \label{fig:exp_end2end_fasttext}
        \includegraphics[width=0.25\textwidth]{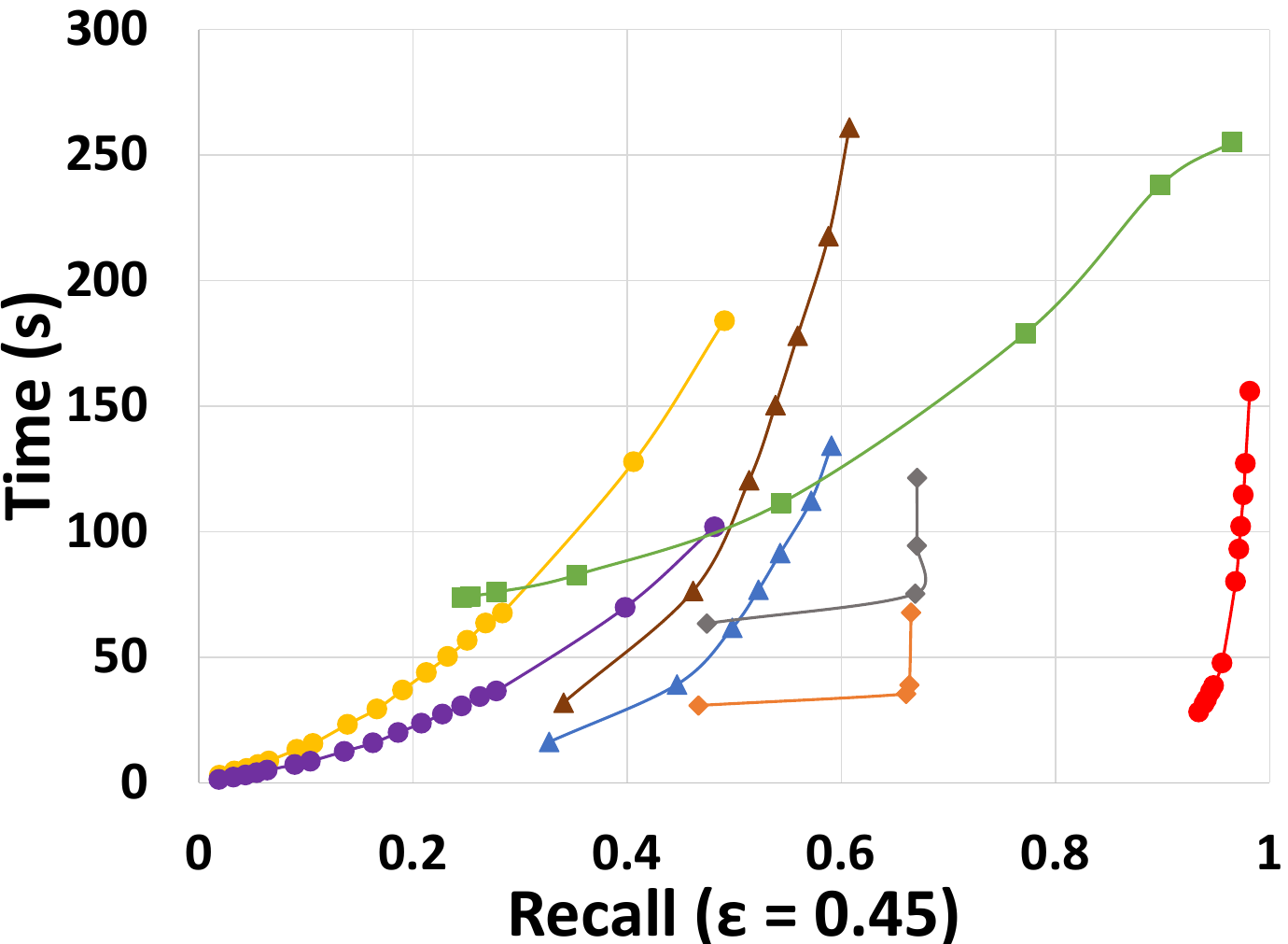}
    }}%
    \qquad
    \subfloat[\centering Word2vec]{{
        \label{fig:exp_end2end_glove}
       \includegraphics[width=0.25\textwidth]{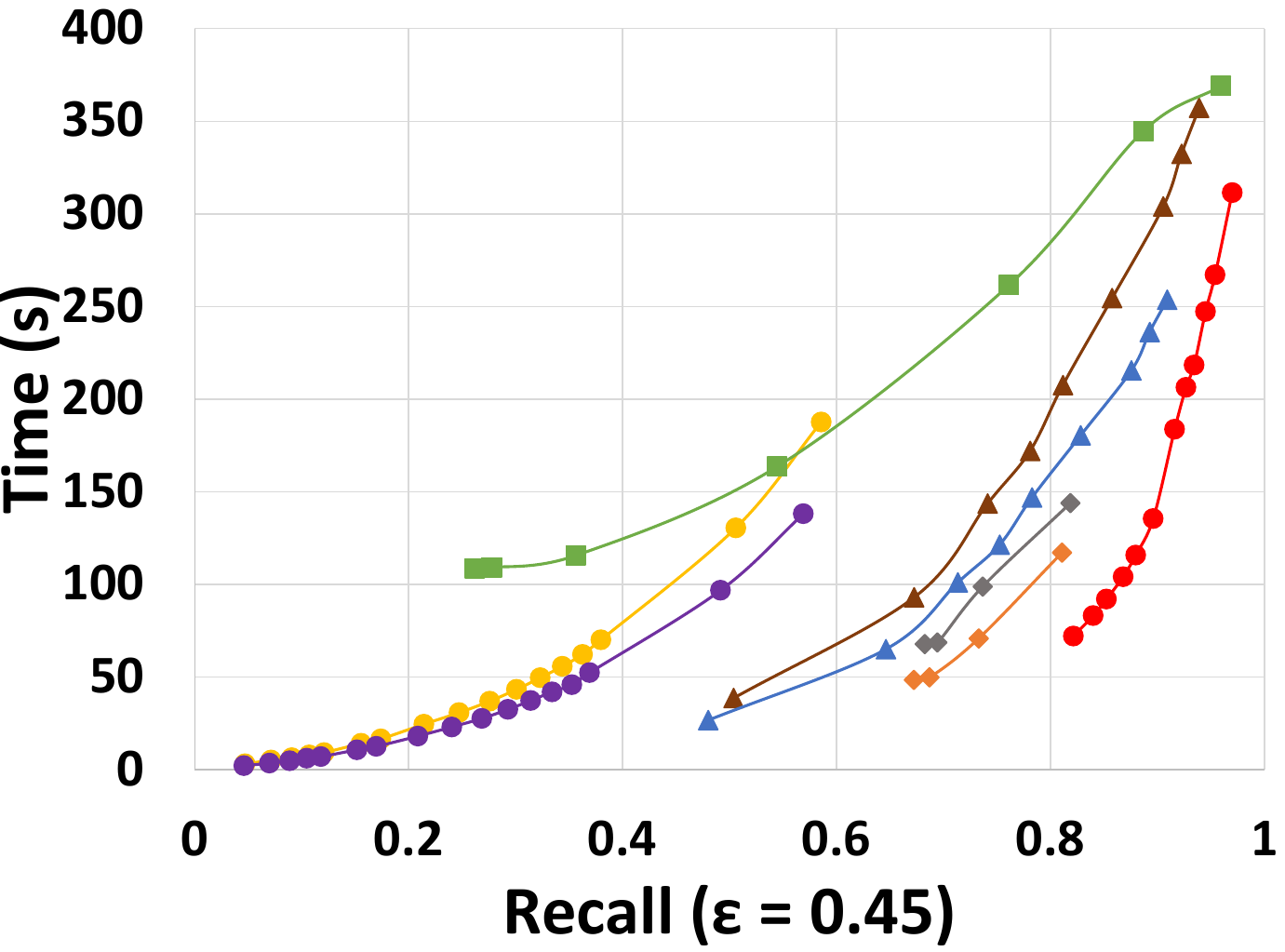}
    }}% 
    \qquad
    \subfloat[\centering Gist]{{
        \label{fig:exp_end2end_fasttext}
        \includegraphics[width=0.25\textwidth]{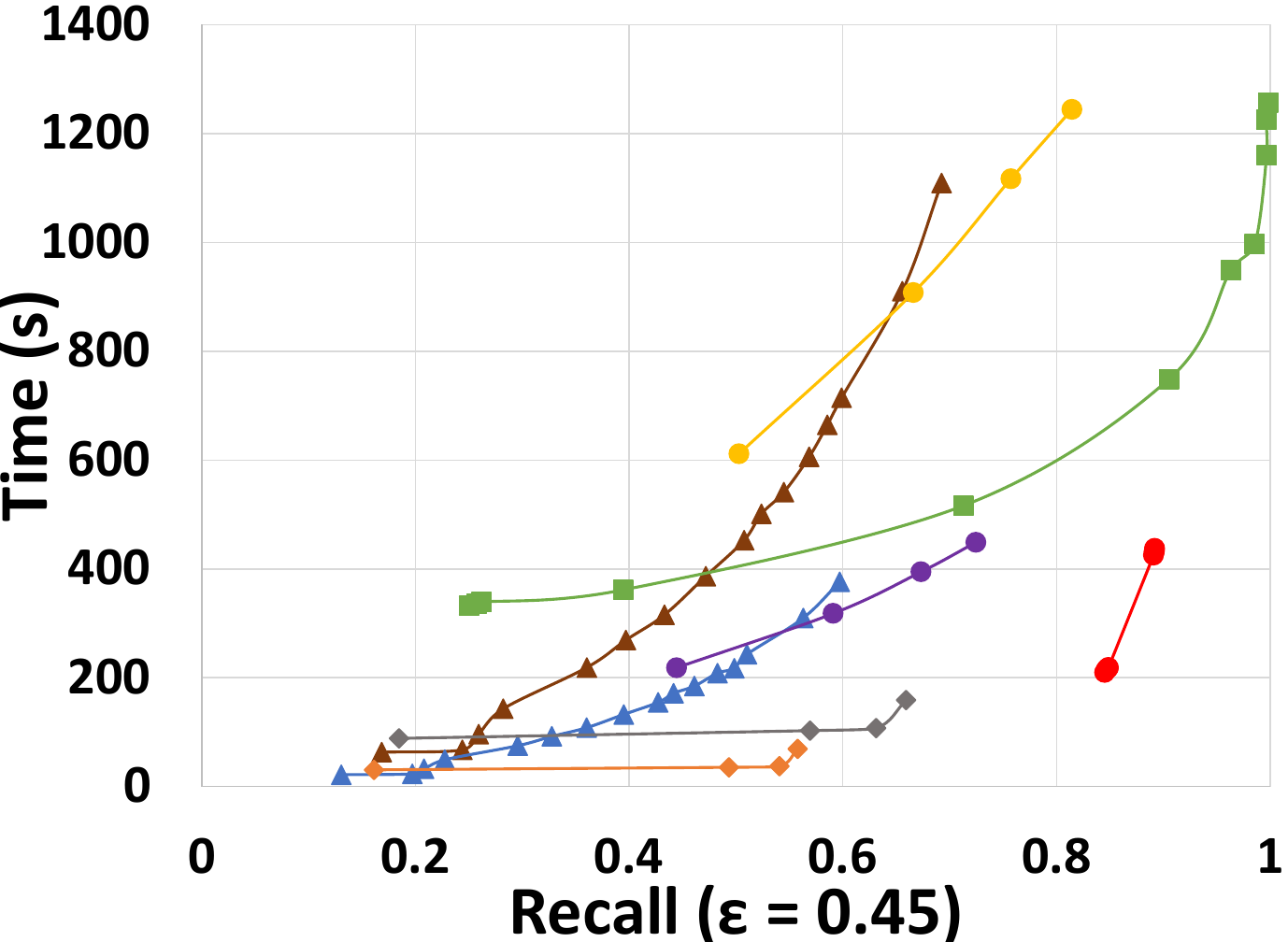}
    }}%
    \caption{Speed-quality trade-off curves for XJoin, the approximate methods and their Xling-enhanced versions on the selected datasets and $\epsilon$, and other cases are also similar.}%
    \label{fig:tradeoff-exps}%
\end{figure*}

In this evaluation, We fix the dataset and $\epsilon$, then vary some key parameters of the approximate baselines and XJoin, i.e., $n_p$ in LSH, the branching factor and the ratio $\rho$ in KmeansTree, $W$ and bit array length in Naive-LSBF, $C$ and $p$ in IVFPQ, and the XDT selection mode (mean or FPR based) and $\tau$ in XJoin. For LSH and KmeansTree, we always set the parameters of their Xling-enhanced versions the same as the original version, in order to make the comparison between them fair. The varied parameters result in varied end-to-end similarity join time and quality, based on which we get the speed-quality trade-off curves, as illustrated in Figure~\ref{fig:tradeoff-exps}.

We select three cases to present in Figure~\ref{fig:tradeoff-exps}: the datasets Glove, Word2vec and Gist with $\epsilon = 0.45$, and other cases are similar to them. According to the results, we conclude that (1) XJoin has better trade-off capability than the original version of the approximate baselines, i.e., LSH, IVFPQ, KmeansTree and Naive-LSBF, meaning that XJoin can achieve high quality with minimum processing time, or sacrifice tiny quality for significant efficiency gain. (2) The Xling-enhanced versions of the approximate baselines (i.e., LSH-Xling, IVFPQ-Xling and KmeansTree-Xling) have significant better trade-off capability than the original versions, which also means Xling successfully accelerates the original versions with a relatively negligible quality loss. This demonstrates the generality and usefulness of Xling to further enhance the existing similarity join methods and make them more practical.


\subsection{Generalization evaluation}
\label{sec:exp-generalization}
In this section we evaluate the generalization capability of Xling and XJoin. Another 150k dataset is sampled for each original dataset. We call the first 150k datasets used in previous experiments ``the first 150k'' or add ``-1st'' after the dataset name, while call the new dataset ``the second 150k'' or add ``-2nd'' after the name. The first and second 150k have no overlap, except NUS-WIDE. 

\begin{figure*}[h]%
    \centering
    \subfloat{{
        \label{fig:legend_generalization}
        \includegraphics[width=1\textwidth]{figures/crop_legend_tradeoff.pdf}
    }}%
    \qquad
    \subfloat[\centering Glove-2nd]{{
        \label{fig:exp_generalization_glove}
        \includegraphics[width=0.25\textwidth]{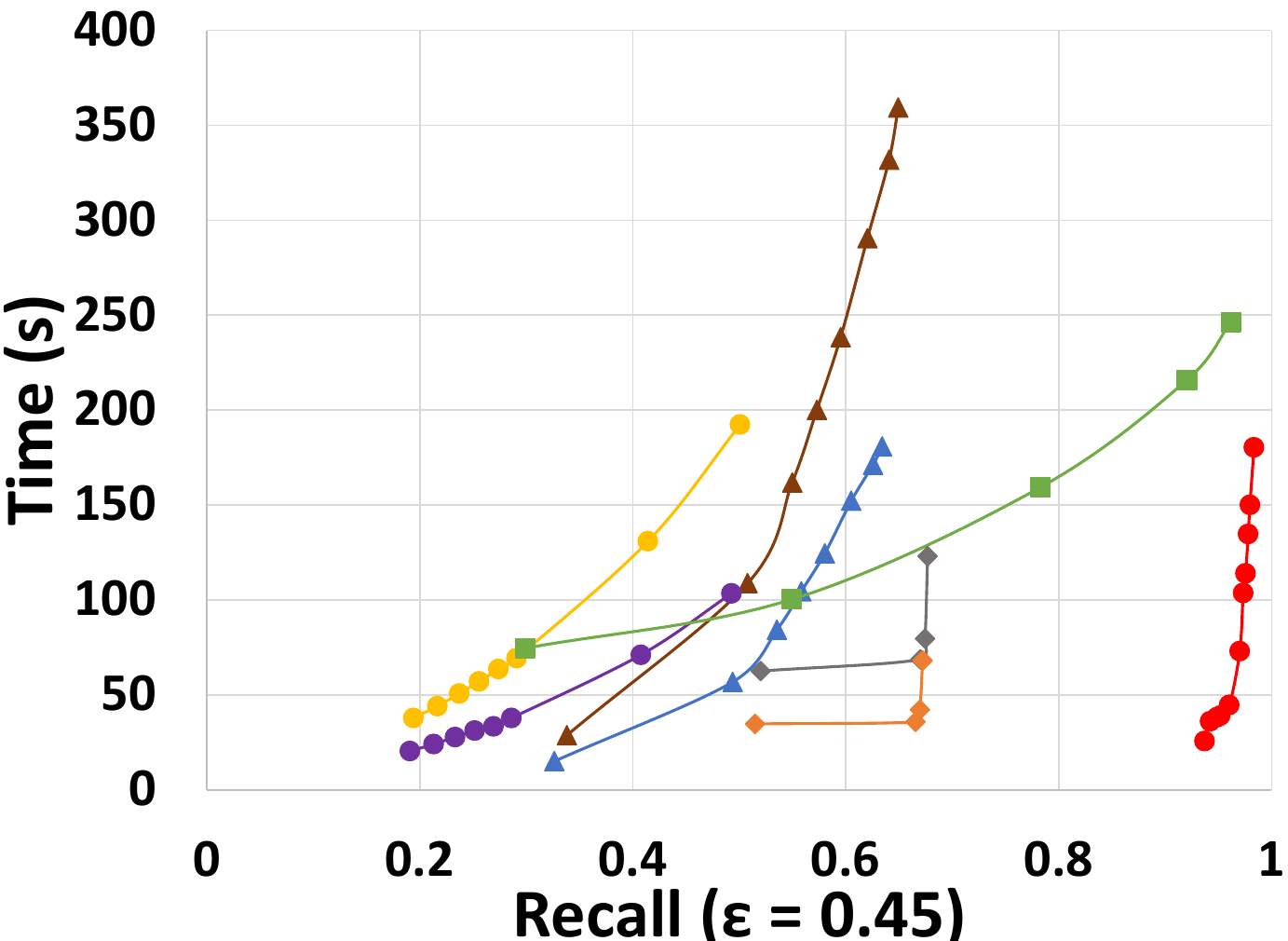}
    }}%
    \qquad
    \subfloat[\centering Word2vec-2nd]{{
        \label{fig:exp_generalization_word2vec}
       \includegraphics[width=0.25\textwidth]{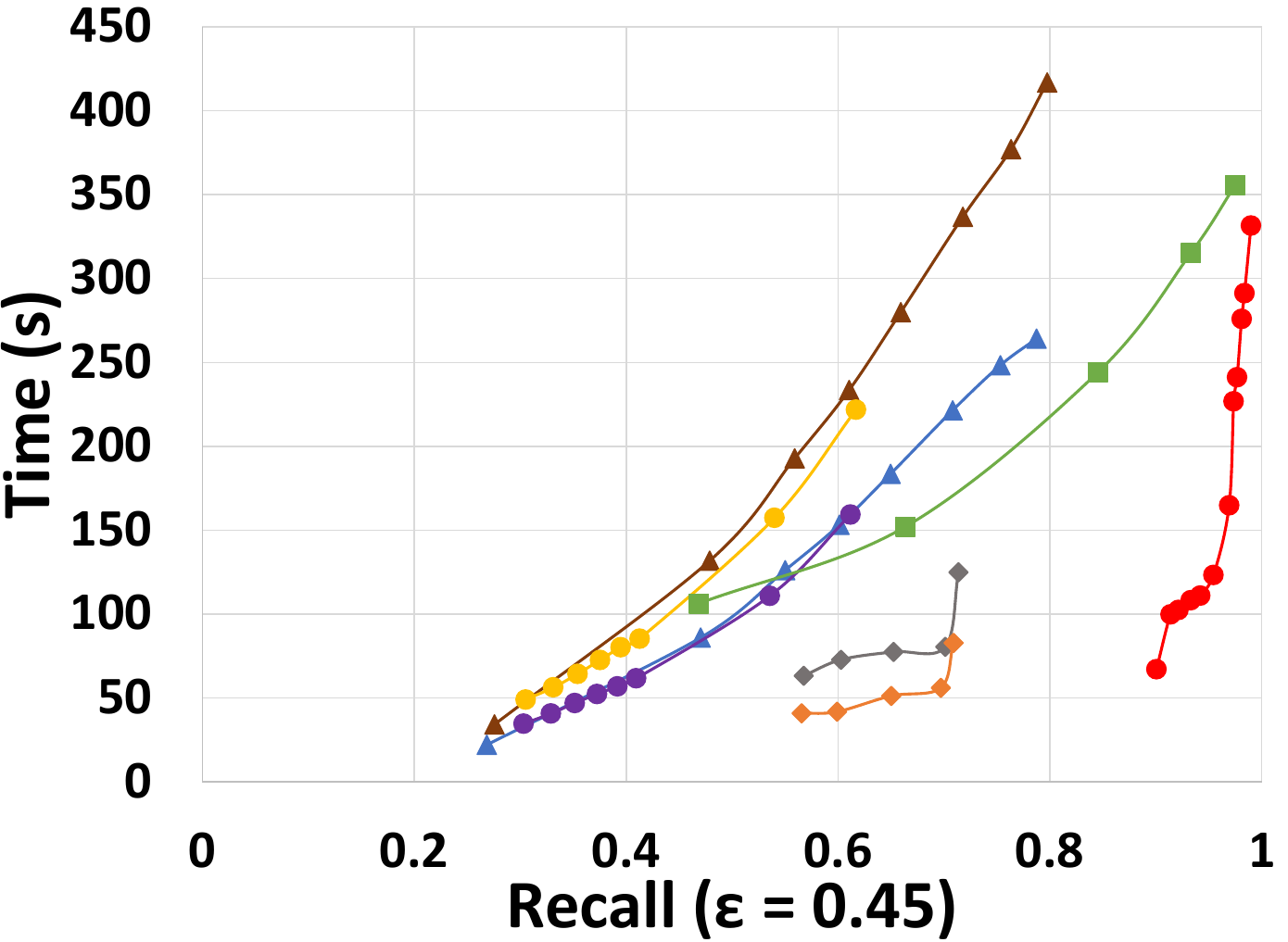}
    }}% 
    \qquad
    \subfloat[\centering Gist-2nd]{{
        \label{fig:exp_generalization_gist}
        \includegraphics[width=0.25\textwidth]{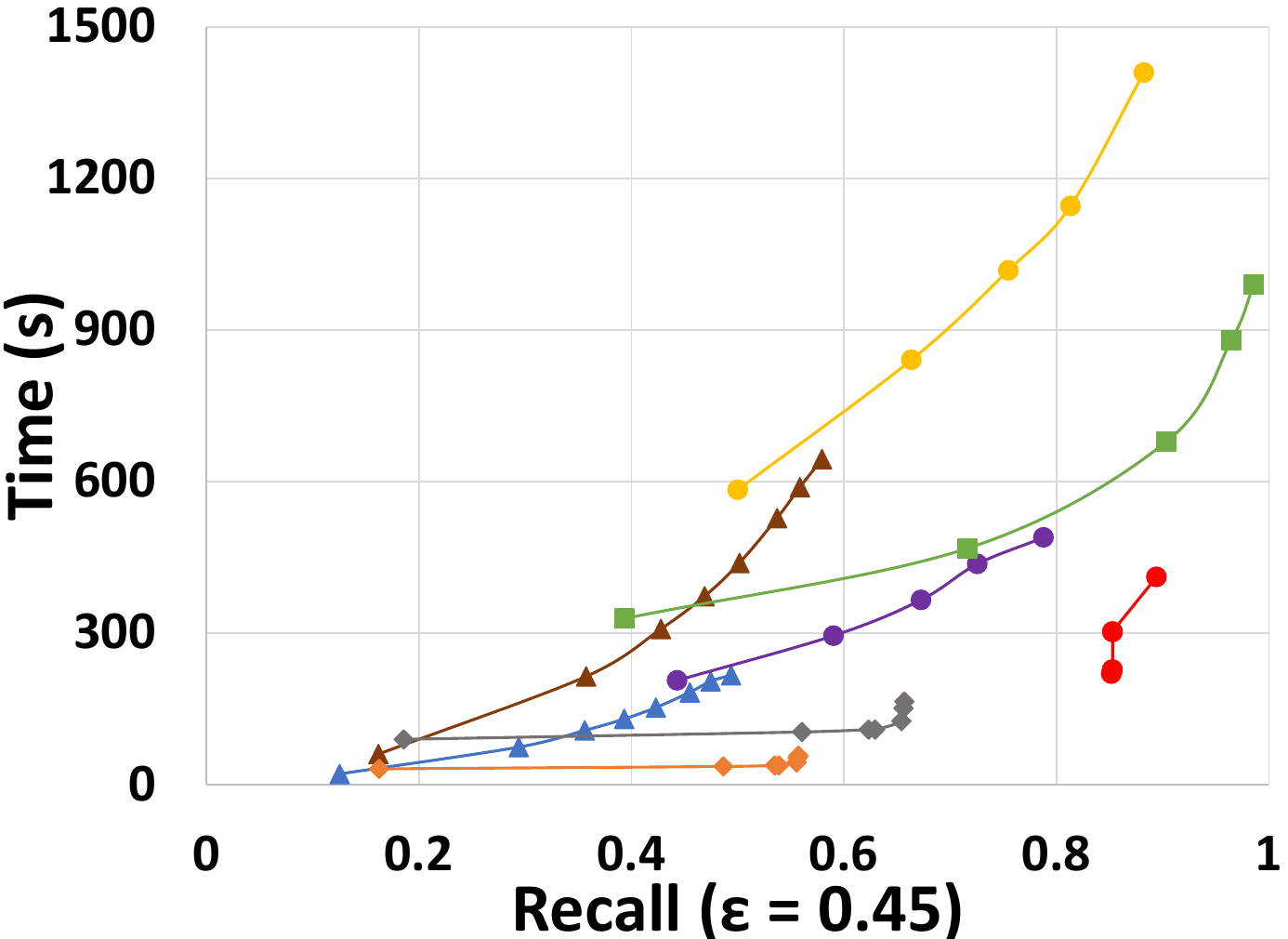}
    }}%
    \caption{Speed-quality trade-off curves for XJoin, the approximate methods and their Xling-enhanced versions on the second 150k datasets, where all Xlings are pre-trained on the original 150k dataset without re-training for the second}%
    \label{fig:generalization-exps}%
    
    \subfloat{{
        \label{fig:legend_robustness_exp}
        \includegraphics[width=1\textwidth]{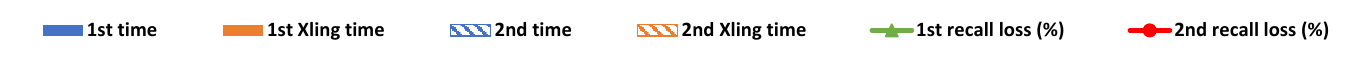}
    }}%
    \qquad
    \subfloat[\centering Glove 1st vs 2nd]{{
        \label{fig:exp_robust_glove}
        \includegraphics[width=0.25\textwidth]{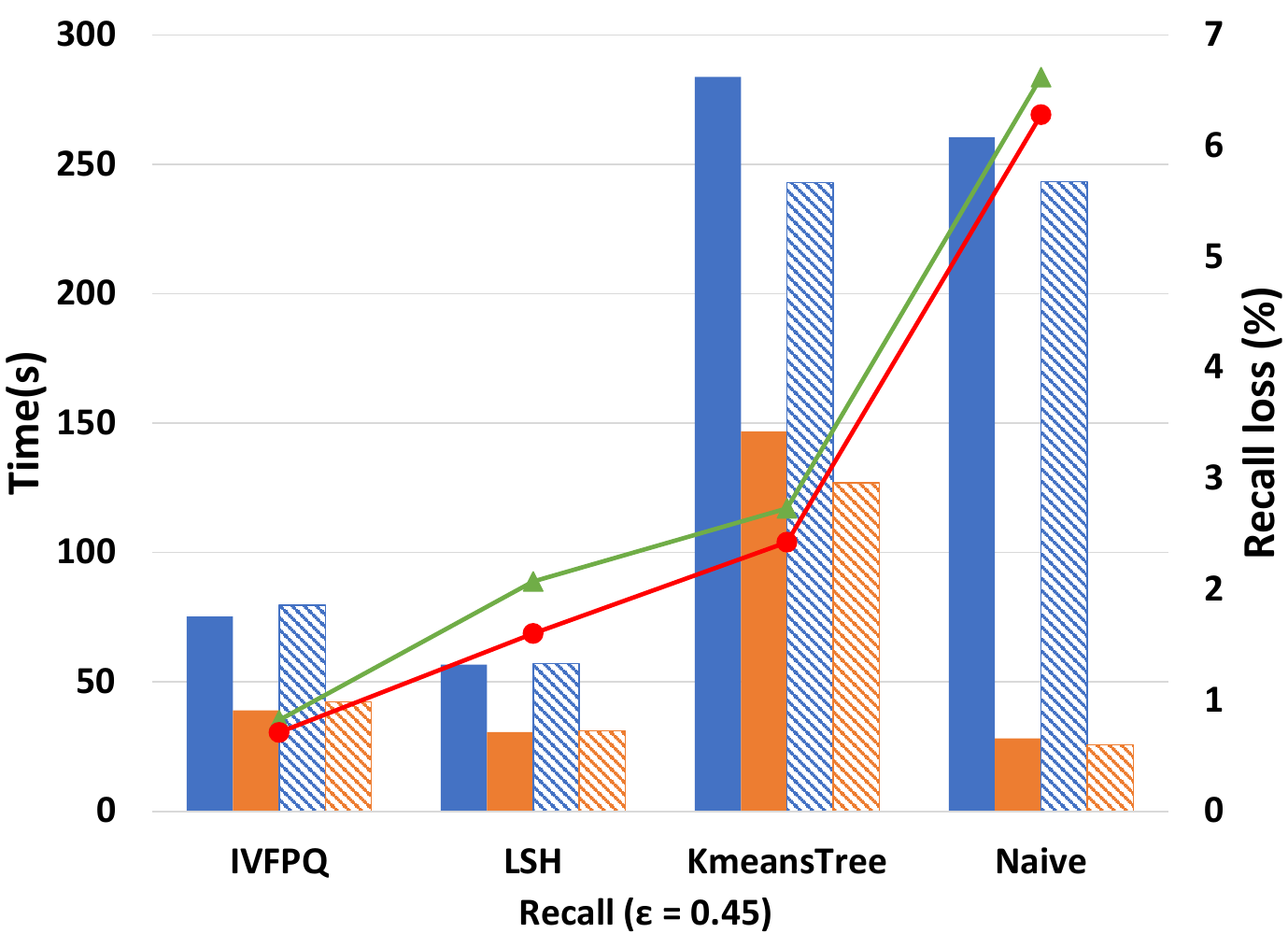}
    }}%
    \qquad
    \subfloat[\centering Word2vec 1st vs 2nd]{{
        \label{fig:exp_robust_word2vec}
       \includegraphics[width=0.25\textwidth]{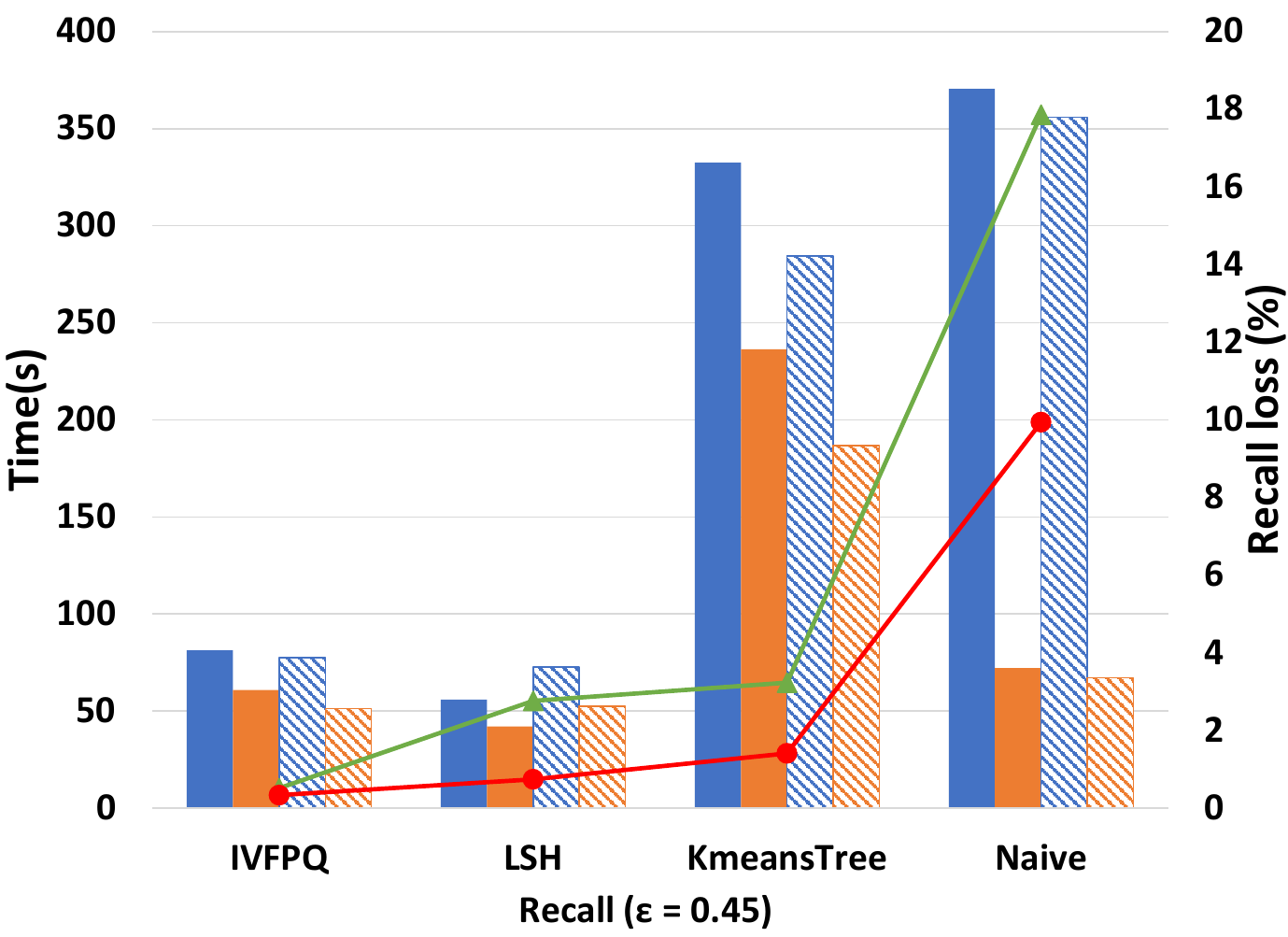}
    }}%
    \qquad
    \subfloat[\centering Gist 1st vs 2nd]{{
        \label{fig:exp_robust_gist}
        \includegraphics[width=0.25\textwidth]{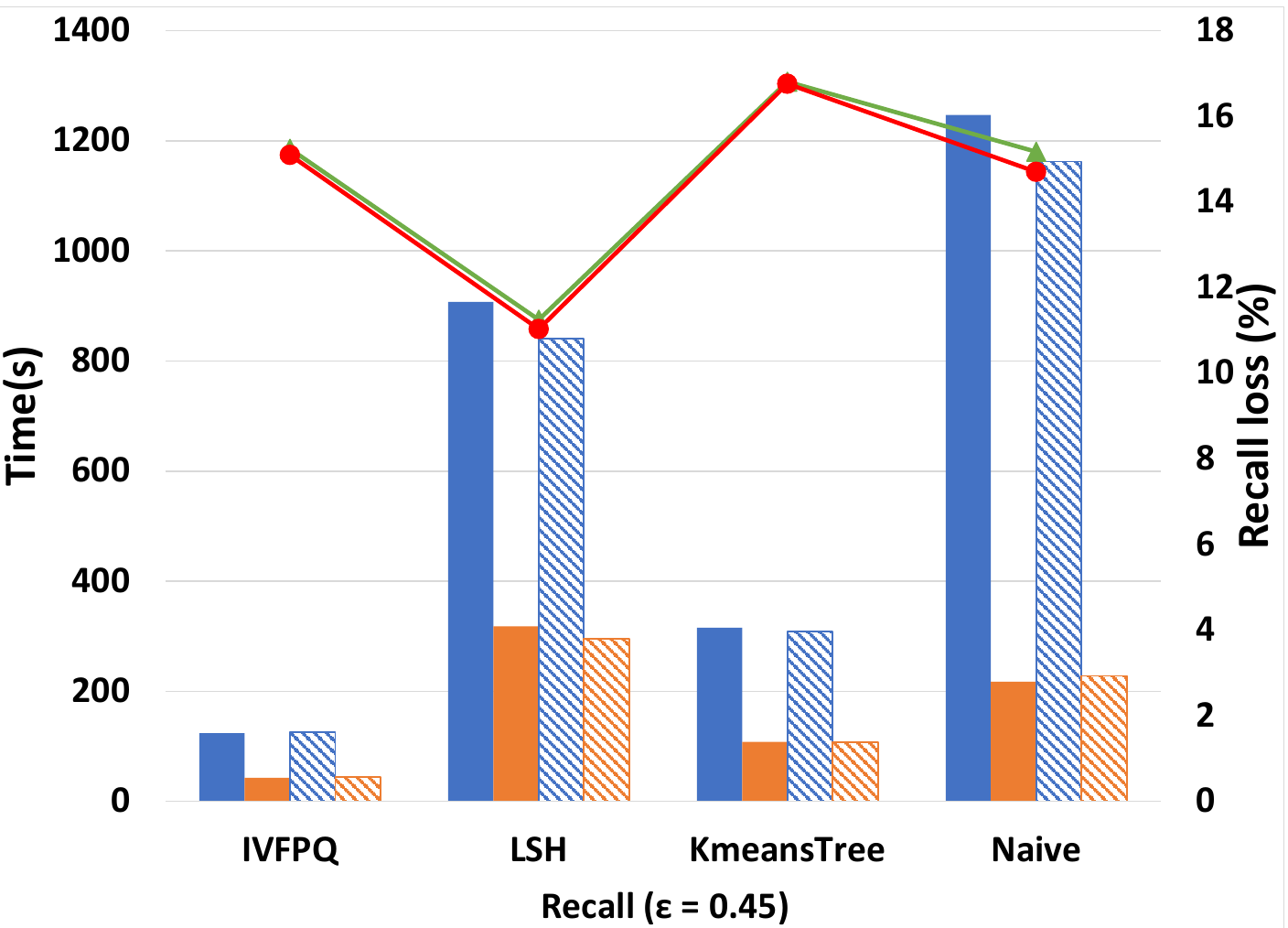}
    }}%
    \caption{The differences of acceleration and recall loss resulting from Xling on the first and second 150k datasets}%
    \label{fig:robust-exps}%
\end{figure*}

We first evaluate the trade-off capabilities of XJoin and Xling-enhanced methods again on the second 150k. All Xlings are those trained on the first 150k and used in the previous experiments. We do not re-train them anymore for the second 150k. As shown in Figure~\ref{fig:generalization-exps}, all those methods present a similar performance and trends as in Figure~\ref{fig:tradeoff-exps}, which prove our statement in the Introduction that Xling and XJoin have great generalization capability thanks to the learning model, and therefore it is not necessary to re-train Xling when the data is updated or even replaced with a fully new dataset, as long as the new data has similar distribution to the old.

To have a further quantitative view about the generalization, we also compare the speed improvement and the recall loss made by Xling when attaching it to the base similarity join methods. In Figure~\ref{fig:robust-exps}, within each method (marked as ``IVFPQ'', ``LSH'', etc.), the first bar (solid and blue, ``1st time'') and second bar (solid and orange, ``1st Xling time'') are the end-to-end running time of the original method and its Xling-enhanced version on the first 150k, while the third bar (diagonal lines and blue, ``2nd time'') and fourth bar (diagonal lines and orange, ``2nd Xling time'') are the two running time on the second 150k. The green and red lines are the percentage recall loss respectively on the first and second 150k, i.e., the difference between recall of original method and enhanced version over original recall on each dataset. The results show that neither time improvement nor recall loss has a significant difference between first and second 150k, which further prove our methods have outstanding generalization, meaning that they are practical in real world.      

\section{Conclusion}
In this paper we propose Xling, a generic framework of learned metric space Bloom filters for speeding up similarity join with quality guarantee based on machine learning. 
Based on Xling we develop an efficient and effective similarity join method that outperforms the state-of-the-art methods on both speed and quality, as well as having a better speed-quality trade-off capability and generalization capability. We also apply Xling onto those state-of-the-art methods to significantly further enhance them.  Xling has shown the great potential in effectively speeding up a wide range of existing similarity join methods.

\bibliographystyle{IEEEtran}
\bibliography{IEEEabrv,main}

\end{document}